\documentclass[11pt]{article}
\pdfoutput=1
\usepackage{booktabs} % To thicken table lines

\usepackage[normalem]{ulem}
\usepackage{float}
\usepackage{epsfig}
\usepackage{cite}
\usepackage{amsmath}
\usepackage[footnotesize]{caption}
\usepackage{slashed}
\usepackage{xcolor}
\usepackage[utf8]{inputenc}
\usepackage{tikz}
\usetikzlibrary{shapes.geometric}
\usetikzlibrary{arrows.meta,arrows}
\usetikzlibrary{quotes,angles}

\usepackage{multicol}
\usepackage{multirow}
\usepackage{array}
\numberwithin{equation}{section}

\newcommand{\be}{\begin{equation}}
\newcommand{\ee}{\end{equation}}
\newcommand{\beq}{\begin{eqnarray}}
\newcommand{\eeq}{\end{eqnarray}}

\usepackage{bm}

\begin{document}

\title{
%\vspace*{-3.7cm}
%\phantom{h} \hfill\mbox{\small KA-TP-10-2017}\\[-1.1cm]
%\vspace*{0.7cm}
%\\[1cm]
%\vspace{13mm}
\textbf{CP-violation, Asymmetries and Interferences in $t \bar{t} \phi$ \\[4mm]}}

\date{}
\author{
Duarte Azevedo$^{1,2\,}$\footnote{E-mail:
  \texttt{duarte.azevedo@kit.edu}} ,
Rodrigo Capucha$^{3\,}$\footnote{E-mail:
\texttt{rscapucha@fc.ul.pt}} ,
Ant\'{o}nio Onofre$^{4\,}$\footnote{E-mail:
\texttt{antonio.onofre@cern.ch}} ,
Rui Santos$^{3,5\,}$\footnote{E-mail:
  \texttt{rasantos@fc.ul.pt}} 
\\[5mm]
{\small\it
$^1$Institute for Theoretical Physics, Karlsruhe Institute of Technology,} \\
{\small\it 76128 Karlsruhe, Germany} \\[3mm]
{\small \it
$^2$Institute for Astroparticle Physics, Karlsruhe Institute of Technology,} \\
{\small \it 76344 Karlsruhe, Germany.} \\[3mm]
{\small\it $^3$Centro de F\'{\i}sica Te\'{o}rica e Computacional,
    Faculdade de Ci\^{e}ncias,} \\
{\small \it    Universidade de Lisboa, Campo Grande, Edif\'{\i}cio C8
  1749-016 Lisboa, Portugal} \\[3mm]
{\small\it
$^4$ Departamento de F\'{\i}sica , Universidade do Minho, 4710-057 Braga, Portugal} \\[3mm]
{\small\it
$^5$ISEL -
 Instituto Superior de Engenharia de Lisboa,} \\
{\small \it   Instituto Polit\'ecnico de Lisboa
 1959-007 Lisboa, Portugal} \\
}

\maketitle

\begin{abstract}
\noindent

In this paper, we use the associated production of top-quark pairs ($t\bar{t}$) with a generic scalar boson ($\phi$) at the LHC ($pp\rightarrow t\bar{t}\phi$) to explore the sensitivity of a large set of observables to the sign of the CP mixing angle ($\alpha$), 
present in the coupling between the scalar boson and the top quarks. The mass of the scalar boson is set to $m_{\phi}=125$~GeV (the Standard Model Higgs boson mass) and its coupling to top-quarks is varied such that $\alpha=$ 0$^\circ$, 22.5$^\circ$,  45.0$^\circ$, 67.5$^\circ$, 90.0$^\circ$, 135.0$^\circ$ and 180.0$^\circ$. Dileptonic final states of the $t\bar{t}\phi$ system are used ($pp\rightarrow b\ell^+\nu_\ell\bar{b}\ell^-\bar{\nu_\ell}b\bar{b}$), where the scalar boson is expected to decay according to $\phi\rightarrow b\bar{b}$. A new method to reconstruct the scalar mass, originally designed for the low mass regime is used, improving the resolution of the Higgs mass by roughly a factor of two. A full phenomenological analysis is performed using Standard Model (SM) background and signal events generated with \texttt{MadGraph5\_aMC@NLO}, in turn reconstructed using a kinematical fit. The most sensitive CP-observables are selected to compute Confidence Level (CL) limits as a function of the sign of the top quark Yukawa couplings to the $\phi$ boson. We also explore the sensitivity to interference terms using differential distributions and angular asymmetries. Given the significant difference between the pure scalar ($\sigma_{0^+}$) and pure pseudo-scalar ($\sigma_{0^-}$) production cross section values, it is unlikely the $t\bar{t}\phi$ channel alone will be sensitive to the sign of the CP-mixing angle or interference terms, even at the end of the LHC. Using the $b_2^{t\bar{t}\phi}$ and $b_4^{t\bar{t}\phi}$ variables, 
exclusion limits at 95\% CL for the CP-even and CP-odd components of the top quark Yukawa couplings are expected to be set to $\tilde{\kappa} \in$ [-0.698,+0.698] and $|\kappa| \in$ [0.878,1.04], respectively, at the end of the High Luminosity phase of the LHC (HL-LHC) by using the dileptonic decay channel alone.

\end{abstract}

\thispagestyle{empty}
\vfill
\newpage
\setcounter{page}{1}

\section{Introduction}
\hspace{\parindent} %forca identacao

As first discussed by Sakharov~\cite{Sakharov:1967dj}, new sources of CP-violation are needed to explain the matter anti-matter asymmetry observed in the Universe. The CP-nature of the 125 GeV Higgs boson 
needs to be scrutinised and the study of its couplings to fermions and gauge bosons at the Large Hadron Collider (LHC) and future colliders is of the utmost importance. The observation of any deviation from the CP number predicted
by the Standard Model (SM) would open doors to extended scalar sectors which easily provide new sources of CP-violation. Therefore, the search for Beyond the Standard Model (BSM) physics, particularly in the Higgs Yukawa couplings, should be a primary target of the next LHC run. 
  
It is ever more likely that the Higgs boson discovered 10 years ago~\cite{Aad:2012tfa, Chatrchyan:2012ufa}  by the ATLAS and CMS collaborations has couplings to the remaining SM particles that resemble very much the ones predicted by the SM.
It was already settled experimentally that the 125 GeV Higgs is not a pure pseudoscalar state with a 99\% confidence level (CL). However, it is also well-known that a significant CP-odd component is still possible even
in very simple extensions of the SM as is the case of the the CP-violating version of the two-Higgs doublet model (2HDM). Usually referred to as C2HDM~\cite{Lee:1973iz, Ginzburg:2002wt, Khater:2003wq, ElKaffas:2007rq, Grzadkowski:2009iz, Arhrib:2010ju, Barroso:2012wz, Inoue:2014nva,Cheung:2014oaa, Fontes:2014xva, Fontes:2015mea, Chen:2015gaa, Muhlleitner:2017dkd, Fontes:2017zfn}, it has been used extensively as a benchmark model in many phenomenological studies. In particular,
it was shown that in the C2HDM the  CP-odd component of the Yukawa couplings can be varied independently for up-type quarks, down-type quarks and leptons~\cite{Fontes:2015mea}. This in turn means that a dedicated study for each fermion type is needed.
The ratio between the CP-even and the CP-odd component of the Yukawa couplings can in principle be probed at tree-level both in the production and decays of these scalars. However, the proposals that were actually turned into experimental analysis
are only the ones for the top quark~\cite{Gunion:1996xu, Boudjema:2015nda, Santos:2015dja, AmorDosSantos:2017ayi}, in the $\bar t t \phi$ production channel and for the tau lepton~\cite{Berge:2008wi, Berge:2008dr, Berge:2011ij, Berge:2014sra, Berge:2015nua} in the decay channel.
ATLAS and CMS have studied the CP-nature of the 125~GeV Higgs boson in these two channels. With the two photon final state channel of a Higgs boson produced in association with top quarks, 
$pp \to \bar t t (H\rightarrow \gamma\gamma)$, ATLAS and CMS~\cite{Sirunyan:2020sum, Aad_2020} excluded the pure CP-odd hypothesis at 3.9$\sigma$ and obtained a 95\% CL observed (expected) 
exclusion upper limit for the CP mixing angle of 43$^{\circ}$~(63$^{\circ}$).  The first measurement of the tau lepton CP mixing angle was performed by CMS~\cite{CMS-PAS-HIG-20-006} using $\sqrt{s} =$ 13~TeV data with
an integrated luminosity of 137~fb$^{-1}$.  The measurement yielded a CP mixing angle of 4$^{\circ}$ $\pm$ 17$^{\circ}$, and an observed (expected) exclusion upper limit of 36$^{\circ}$~(55$^{\circ}$).

Besides the top-quark and tau-lepton  Yukawa couplings, the only theoretical studies available for the remaining fermions are for the b-quark. The direct observation via the production mode $pp \to \bar b b \phi$ 
was discussed in~\cite{Ghosh:2019dmv, Azevedo:2020vfw} showing that it will be extremely hard to repeat what was done successfully for the top quark. As for the b-quark decays, the authors of~\cite{Alonso:2021boj} 
discussed the prospects for probing the CP structure in the $\bar b b \phi$ and $\bar c c \phi$ vertices by measuring the heavy-quark polarization from the hadronization to the lightest flavoured baryons ($\Lambda_q$), which 
preserves  the original quark spin in the infinite-mass limit because it decays weakly. They showed that only at a future muon collider would we have some chance of having sensitivity to those vertices.  The study
of the $\bar b b \phi$ using kinematic shapes was performed in~\cite{Grojean:2020ech}. Note that contrary to CP-violation in the Yukawa couplings which is a tree-level process, even for a simple model like the C2HDM, CP-violation in the gauge couplings is a higher order process that has to be measured indirectly. Just considering as an example the most general $W^+ W^- \phi$ vertex it was shown in~\cite{Huang:2020zde}, using the C2HDM as benchmark, that even at the high luminosity stage of
the LHC it will be very unlikely to have sensitivity to the CP-violating operator of the  $W^+ W^- \phi$ vertex.

The different possible approaches for probing CP-violating couplings from loop processes involving gauge bosons was recently discussed in~\cite{Haber:2022gsn}. It was also 
examined in detail the origing of CP violation in the Lagrangian:  Yukawa couplings stems from P-violation while CP violation coming from C violation has its origin in the scalar sector. In the latter scenario, probing CP violation would imply resourcing to a combination of three decays that would only make sense if new scalars were found. Finally, very recently it was shown in~\cite{Hermann:2022vit} that attention must be paid to  both NLO corrections and off-shell effects because they play an important role 
in the observables that are used to determine the CP nature of the Higgs in its Yukawa couplings.

In this paper, we use the associated production of top-quark pairs with a generic scalar boson at the LHC ($pp\rightarrow t\bar{t}\phi$) to explore the sensitivity of a large set of observables to the sign of the CP-odd component (reflected on the values of the mixing angle $\alpha$), 
present in the coupling between the scalar boson and the top quarks. Over the years we have not only proposed a number of new variables to probe CP-violation in this channel but we have also tested the most relevant ones present in the literature. 
With this knowledge we have now combined these variables in order to extract the best possible limit on the CP-odd component of the scalar. Although previous papers have discussed how to probe the sign of the CP-odd component of the Yukawa coupling via the mixing angle $\alpha$~\cite{Ellis_2014, Gon_alves_2018, Gon_alves_2022} a proper analysis at detector level with the main backgrounds included was never performed. Therefore, we will present for the first time a study that include all available variables in the literature and that also discusses
the interference terms. The Higgs boson mass is set to $m_{\phi}=125$~GeV and dileptonic final states of the $t\bar{t}\phi$ system are used ($pp\rightarrow b\ell^+\nu_\ell\bar{b}\ell^-\bar{\nu_\ell}b\bar{b}$), with $\phi\rightarrow b\bar{b}$. A new method to reconstruct the scalar mass, originally designed for the low mass regime is used, improving the resolution of the Higgs mass by roughly a factor of two~\cite{Azevedo:2020vfw}. A full phenomenological analysis is performed using SM background and signal events generated with \texttt{MadGraph5\_aMC@NLO}, in turn reconstructed using a kinematical fit. 

The paper is organized as follows. The different CP-observables are presented in Section~\ref{sec:TH}. The differential production cross sections for the various CP signals and the interference terms, are discussed in Section~\ref{sec:interference}. The event generation, selection and kinematic reconstruction are presented in Section~\ref{sec:generation} and the results are analysed in Section~\ref{sec:results}. Our main conclusions are presented in Section~\ref{sec:conclusions}.

%%%%%%%%%%%%%%%. Lagrangian
\section{CP-observables \label{sec:TH}}
\hspace{\parindent} %forca identacao

The most general top quark-Higgs interaction can be parameterized as
\begin{equation}
{\cal L} = \kappa_t y_t  \bar t (\cos \alpha + i \gamma_5 \sin \alpha) t \phi \, = y_t \bar t (\kappa + i \gamma_5 \tilde{\kappa}) t \phi ,
\label{eq:higgscharacter}
\end{equation}
where the real parameter $\kappa_t$ describes the magnitude of the coupling strength with respect to the SM and $\alpha$ is the CP-mixing angle. The last part of the previous expression is another parameterization of the same coupling, the mapping between them given by $\kappa = \kappa_t \cos \alpha$ and $\tilde{\kappa} = \kappa_t \sin \alpha$. For the SM hypothesis, i.e. the CP-even case ($\phi = H$), we fix $\kappa_t$=1 and  $\alpha = 0^\circ$. Alternatively, for the CP-odd case ($\phi = A$), we consider $\alpha = 90.0^\circ$. The Yukawa coupling strength of the top quark to the SM-Higgs boson is given by $y_t=\sqrt{2}m_t/v$, proportional to the top quark mass ($m_t$) and the electroweak vacuum expectation value $v$.

Several CP-observables have been proposed in the literature to probe the CP-nature of the top-Higgs couplings at the LHC or future colliders, using mainly the $t\bar{t} \phi$ production channel~\cite{Bernreuther_1994, Gunion:1996xu, Bhupal_Dev_2008, Frederix_2011, Ellis_2014, Khatibi_2014, Demartin_2014, Kobakhidze_2014, Bramante_2014, Boudjema:2015nda, He_2015, Santos:2015dja, Gritsan_2016, Dolan_2016, Gon_alves_2016,  Buckley_2016_v2, Mileo_2016, Buckley_2016, AmorDosSantos:2017ayi, Gon_alves_2018, Azevedo_2018, Li_2018, Ferroglia_2019, Faroughy_2020, Azevedo:2020vfw, Azevedo:2020fdl, Bortolato_2021, new_obs, Gon_alves_2022, Barman_2022}. These observables can be sensitive to the nature of the coupling but also allow discrimination of scalar boson signals from irreducible backgrounds at the LHC. Moreover, it has been shown that some of these observables can be used to explore the CP nature of Higgs bosons with masses ranging from 12~GeV to 500~GeV~\cite{Azevedo:2020vfw, Azevedo:2020fdl}. The vast majority of these variables are only sensitive to the square terms $\kappa^2$ and $\tilde{\kappa}^2$ that appear in the cross section of the interaction described in Eq.~\ref{eq:higgscharacter}, missing the interference term between the CP-even and CP-odd couplings (proportional to $\kappa \tilde{\kappa}$). We define these observables as CP-even, as they are invariant under a CP-transformation and, in particular, they are not sensitive to the relative sign of the CP-phase $\alpha$. 
In this paper, from all variables considered, we show results for only a few most sensitive CP-even variables like the $b_2^{t\bar{t}\phi}$ and $b_4^{t\bar{t}\phi}$ defined in the $t\bar{t}\phi$ centre-of-mass (CM) frame~\cite{Ferroglia_2019} according to
\begin{equation}
b^{t\bar{t}\phi}_2  = ( \vec{p}_{t} \times \hat{k}_z ).( \vec{p}_{\bar{t}} \times \hat{k}_z )/ (|\vec{p}_{t}| .  |\vec{p}_{\bar{t}}|) \ \ \text{and} \ \  b^{t\bar{t}\phi}_4 = (p^z_t . p^z_{\bar{t}}) / (|\vec{p}_{t}| . |\vec{p}_{\bar{t}}| )
\label{equ:b2b4}
\end{equation}
where $\vec{p}_{t(\bar{t})}$ and $p^z_{t(\bar{t})}$,  correspond to the total and $z$-component of the top (anti-top) quark momentum measured in the $t\bar{t}\phi$ centre-of-mass system. The beam line direction defines the unit vector $\hat{k}_z$. It is worth noting that $b_2^{t\bar{t}\phi}$ and $b_4^{t\bar{t}\phi}$ have a natural physics interpretation since they only consider the information of the transverse $p^T_{t(\bar{t})}$ (in $b_2^{t\bar{t}\phi}$) and longitudinal $p^z_{t(\bar{t})}$ (in $b_4^{t\bar{t}\phi}$) components of the $t(\bar{t})$ momentum, measured in the $t\bar{t}\phi$ system, normalized to the $t(\bar{t})$  total momentum. As these components depend on the $t$ and $\bar{t}$ polar angles with respect to the $z$-direction, $\theta_t$ and $\theta_{\bar{t}}$, and on the azimuthal angle difference between the top quarks $\Delta\phi_{t\bar{t}}$, they can simply be expressed as  $b_2= \cos\Delta\phi_{t\bar{t}} \times \sin{\theta_t} \times \sin{\theta_{\bar{t}}}$ and $b_4=\cos{\theta_t} \times \cos{\theta_{\bar{t}}}$, i.e. as a function of angular distributions alone. Furthermore, we will consider two additional variables
\begin{equation}
\sin (\theta^{t\bar{t}\phi}_\phi)*\sin(\theta^{t\bar{t}}_{\bar{t}}), \ \ \ \ \ \ \ \ \ \ \ \ \sin (\theta^{t\bar{t}\phi}_\phi)*\sin(\theta^{\bar{t}}_{\bar{b}_{\bar{t}}}) \ \ \text{(seq. boost)},
\label{equ:sinsin}
\end{equation}
where $\theta^X_Y$ is the angle between the direction of the $Y$ system 3-momentum (in the rest frame of $X$) with respect to the momentum direction of the $X$ system (in the rest frame of its parent system), as defined in~\cite{Santos:2015dja, AmorDosSantos:2017ayi}. Angular asymmetries associated to each one of the observables in  Eqs. \ref{equ:b2b4} and \ref{equ:sinsin}, are defined following~\cite{Santos:2015dja},
\begin{eqnarray}
\ensuremath{A_c[Z]= \frac  {\sigma(Z>x_c)-\sigma(Z<x_c)}{ \sigma(Z>x_c)+\sigma(Z<x_c) } },
\label{equ:asymmetries}
\end{eqnarray}
where $\sigma(Z>x_c)$ and $\sigma(Z<x_c)$ correspond to the total cross section for each observable ($Z$), above and below a specific cut-off value $x_c$, respectively. The cut-off values for each observable are chosen to be the ones where the difference between the respective asymmetries for the CP-even against the CP-odd case are largest. In Tab.~\ref{tab:AsymGen}, the cut-off values ($x_c$) are shown for each observable together with the asymmetry values at next-to-leading order (NLO) with parton shower effects. This is given for the CP-even ($\alpha=0.0^\circ$), CP-odd ($\alpha=90.0^\circ$) and mixed cases without any cuts. For completeness, the dominant background $t\bar{t}b\bar{b}$ is also shown. The $b_2^{t\bar{t}\phi}$ and $b_4^{t\bar{t}\phi}$ variables show the strongest variation of asymmetry values as a function of the mixing angle, hence making these observables particularly sensitive to the absolute value of the mixing angle.

\begin{table}[h]
\renewcommand{\arraystretch}{1.3}
\begin{center}
 \hspace*{-2mm}
  \begin{tabular}{c|c|ccccccc|c}
    \toprule
                &         & \multicolumn{8}{c}{ MadGraph5 @ NLO+Shower (no cuts applied)}        \\[-1mm]
    Asymmetries &  $x_c$  & \multicolumn{7}{c}{ $t\bar{t}\phi$ signal mixing angle $\alpha$ (deg.)} &    \\
                &         &  $0.0^\circ$ & $22.5^\circ$ & $45.0^\circ$ & $67.5^\circ$ & $90.0^\circ$ & $135.0^\circ$ & $180.0^\circ$ &  $t\bar{t}b\bar{b}$ \\
    \midrule 
% NLO+Shower values 
      \hspace*{-3mm}$A_c[b_2^{t\bar{t}\phi}]$  
        & $-0.30$ & $-0.35$ & $-0.31$ & $-0.15$ & $+0.15$ & $+0.34$ & $-0.14$ & $-0.36$	 & $-0.17$   \\
      \hspace*{-3mm}$A_c[b_4^{t\bar{t}\phi}]$ 
        & $-0.50$ & $+0.41$ & $+0.37$ & $+0.22$ & $-0.04$ & $-0.22$ & $+0.22$ & $+0.41$	 & $+0.33$   \\
      \hspace*{-3mm}$A_c[\sin (\theta^{t\bar{t}\phi}_\phi)*\sin(\theta^{t\bar{t}}_{\bar{t}})]$ 	
        & $+0.70$ & $-0.27$ & $-0.26$ & $-0.20$ & $-0.09$ & $-0.03$ & $-0.20$ & $-0.27$	 & $-0.56$   \\
      \hspace*{-3mm}$A_c[\sin (\theta^{t\bar{t}\phi}_\phi)*\sin(\theta^{\bar{t}}_{\bar{b}_{\bar{t}}})$]	
        & $+0.60$ & $+0.05$ & $+0.05$ & $+0.07$ & $+0.09$ & $+0.11$ & $+0.06$ & $+0.05$	 & $-0.38$   \\[-2mm]
      \hspace*{-3mm}\text{ (seq. boost)}	
        &  &  &  &  &  &  &  & 	 &   \\
    \bottomrule
  \end{tabular}
\caption{Asymmetries for the $t\bar{t}\phi$ signal as a function of the  mixing angle $\alpha$, as well as for the dominant background $t\bar{t}b\bar{b}$ at NLO+Shower 
(without any cuts), are shown for several observables. Significant differences between the asymmetries for the pure scalar ($\alpha=0.0^\circ$) and pseudo-scalar ($\alpha=90.0^\circ$) cases are observed for several asymmetries.}
\label{tab:AsymGen}
\end{center}
\end{table}

Contrary to CP-even observables, CP-odd variables are sensitive to the sign of $\alpha$ since the interference term may contribute to the total differential cross section. One can show that the only non-zero contributions to the interference term comes from the totally anti-symmetric tensor product of the form $\epsilon_{\mu\nu\gamma\rho} p^\mu_1 p^\nu_2 p^\gamma_3 p^\rho_4$ (with $\epsilon_{1234} = 1$), where $p_i \, (i = 1, ..., 4)$ represents the four momenta associated with the process~\cite{Mileo_2016}. Furthermore, choosing a reference frame where $p^\mu_1 = (E_1, \vec{0})$, $\epsilon_{\mu\nu\gamma\rho} p^\mu_1 p^\nu_2 p^\gamma_3 p^\rho_4$ reduces to a scalar triple product of the form $E_1 \, \vec{p}_2 \cdot (\vec{p}_3 \times \vec{p}_4)$, allowing us to construct simpler CP-odd observables~\cite{Durieux_2015}. In our paper, we will consider two most sensitive CP-odd observables from previous studies~\cite{Ellis_2014, Gon_alves_2018}, where the tensor products already mentioned are evaluated at the $t\bar{t}$ CM frame, resulting in a single triple product of the form $\vec{p}_t \cdot (\vec{p}_{l^+} \times \vec{p}_{l^-})$. We then define $\Delta \phi_{ll}$ which is the angle between the two lepton momenta projected onto the plane perpendicular to the $t$ direction at the centre-of-mass frame of the $t\bar{t}$ system and
\begin{equation}
 \Delta \phi^{t\bar{t}}_{ll} = \text{sgn} [\hat{p}_t \cdot (\hat{p}_{l^+} \times \hat{p}_{l^-})] \ \text{arccos}[(\hat{p}_{t} \times \hat{p}_{l^+}) \cdot (\hat{p}_{t} \times \hat{p}_{l^-})] \ .
\label{equ:deltaphi}
\end{equation}
Both variables are defined in the $[-\pi, \pi]$ range. Notice however that there is a relative minus sign between these two definitions since the sign of $\Delta \phi_{ll}$ is defined as the sign of $\hat{p}_t \cdot (\hat{p}_{l^-} \times \hat{p}_{l^+})$. For additional examples of CP-odd observables see \cite{Bernreuther_1994, Boudjema:2015nda, Mileo_2016, Faroughy_2020, Bortolato_2021, Gon_alves_2022, Barman_2022}.

%%%%%%%%%%%%%%%. Event Generation and Kinematic Reconstruction
\section{Differential Cross Sections and Interference Term  \label{sec:interference}}
\hspace{\parindent} %forca identacao

\begin{figure}[H]
	\begin{center}
		\includegraphics[height = 7.0cm]{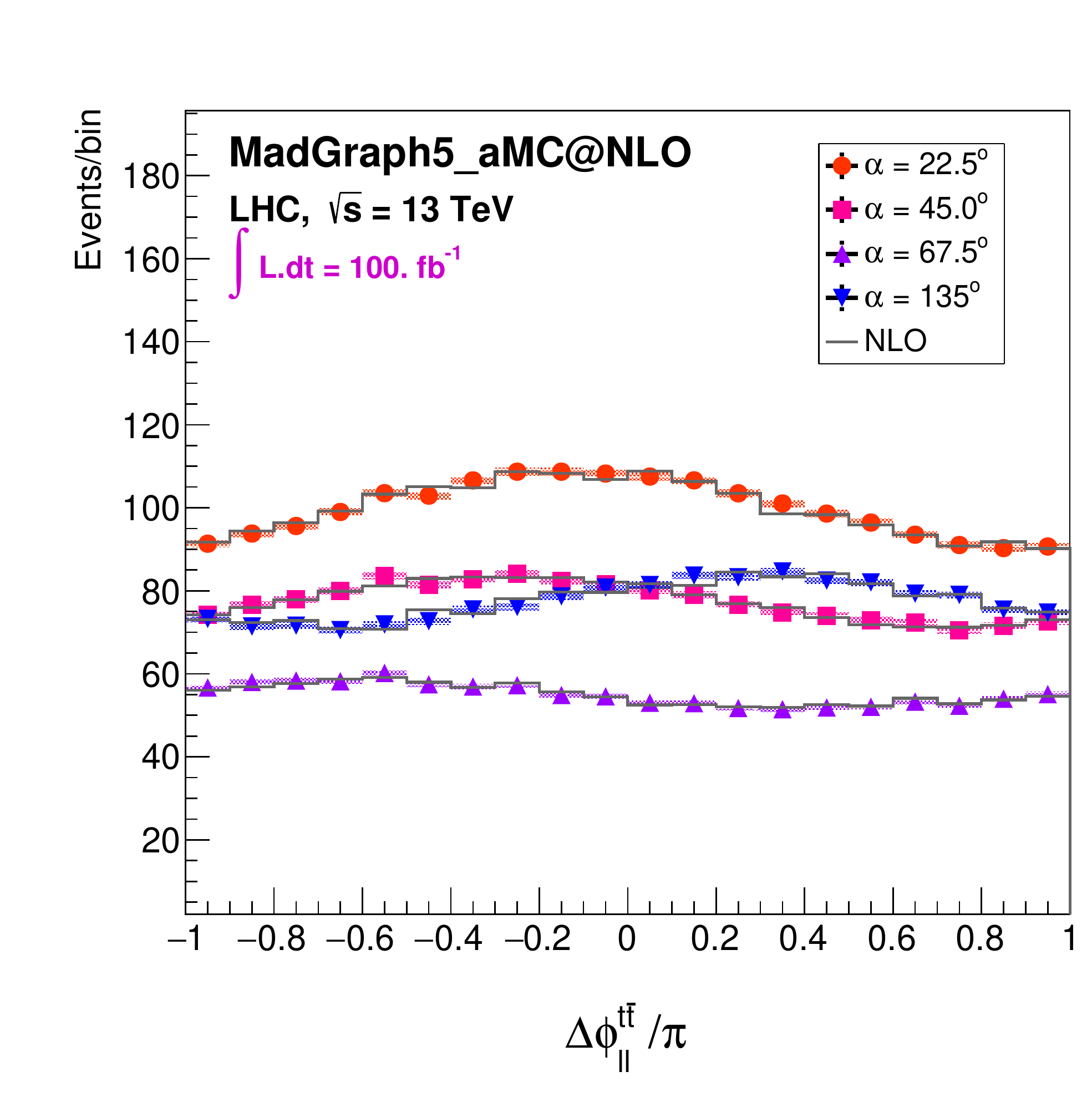}
		\includegraphics[height = 7.0cm]{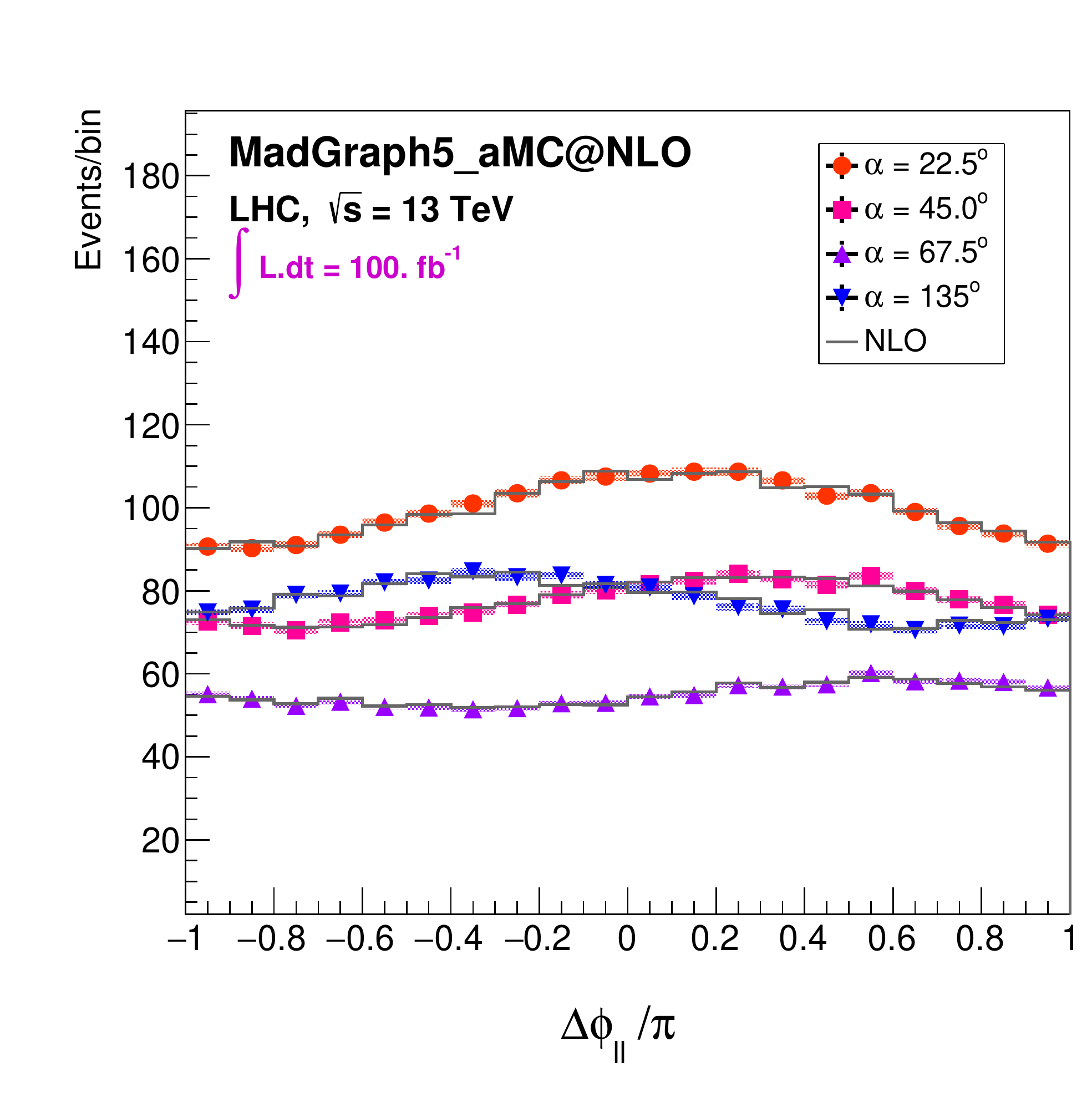} \\
		\includegraphics[height = 7.0cm]{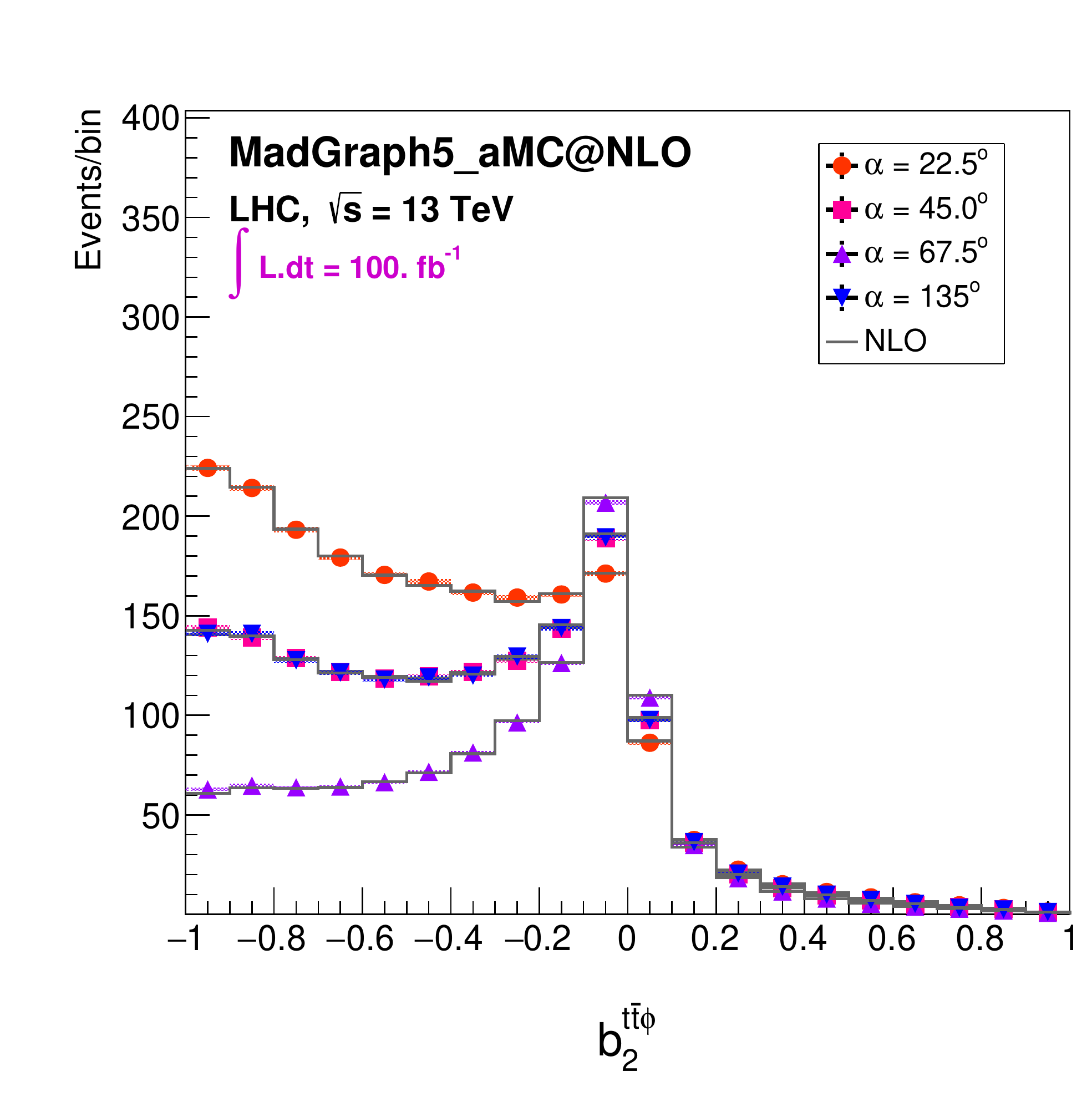}
		\includegraphics[height = 7.0cm]{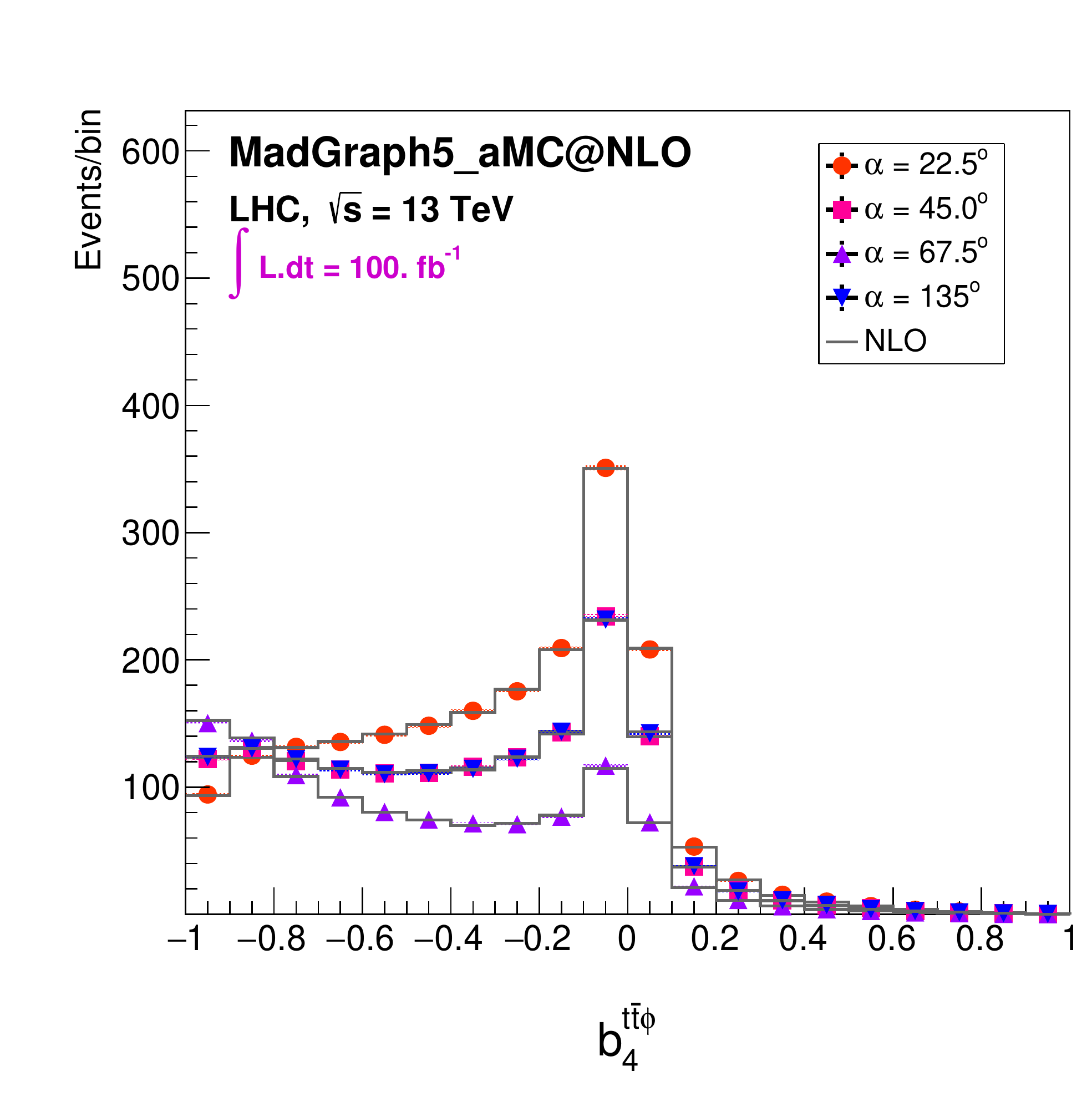} \\
		\includegraphics[height = 7.0cm]{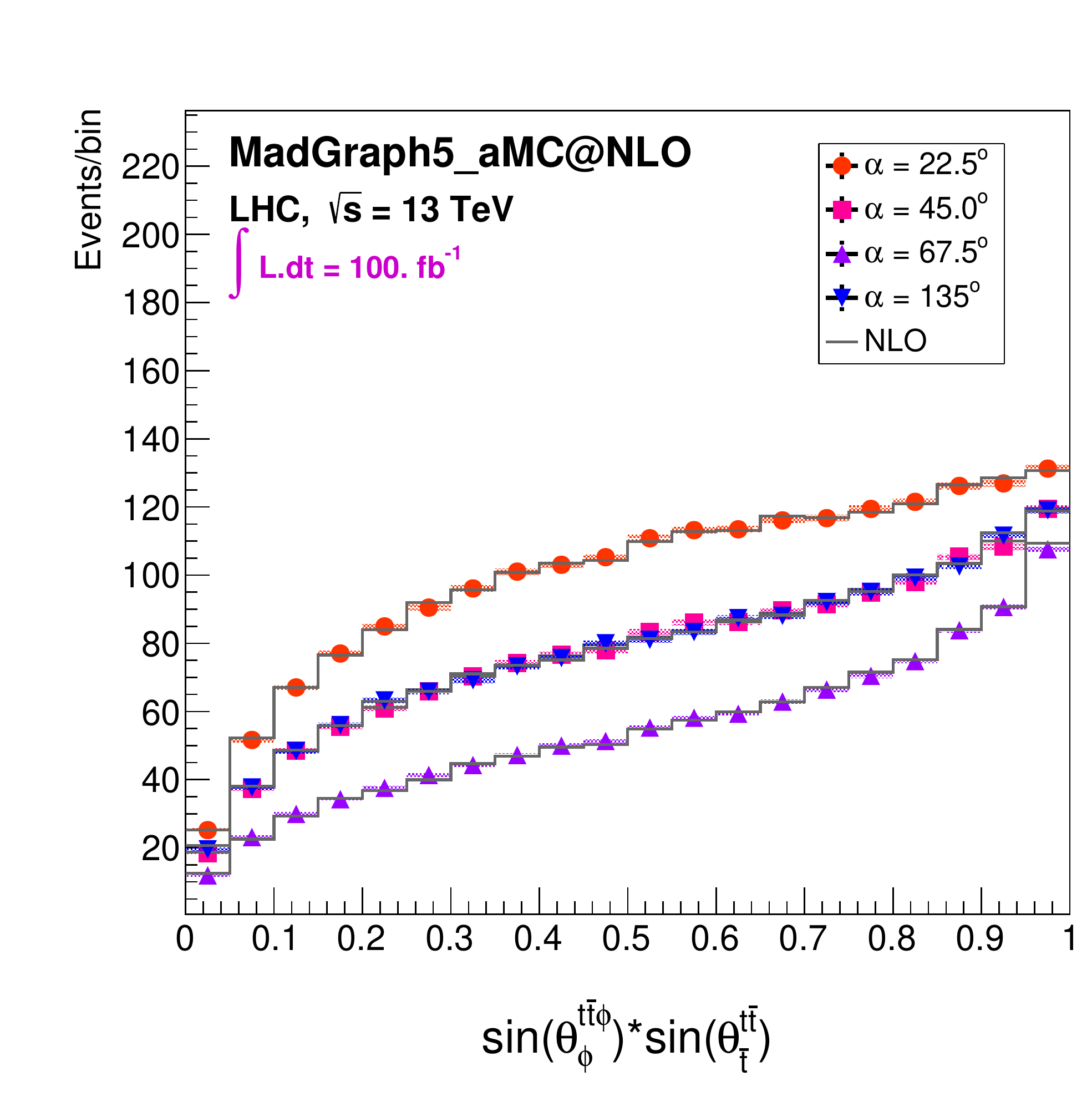}
		\includegraphics[height = 7.0cm]{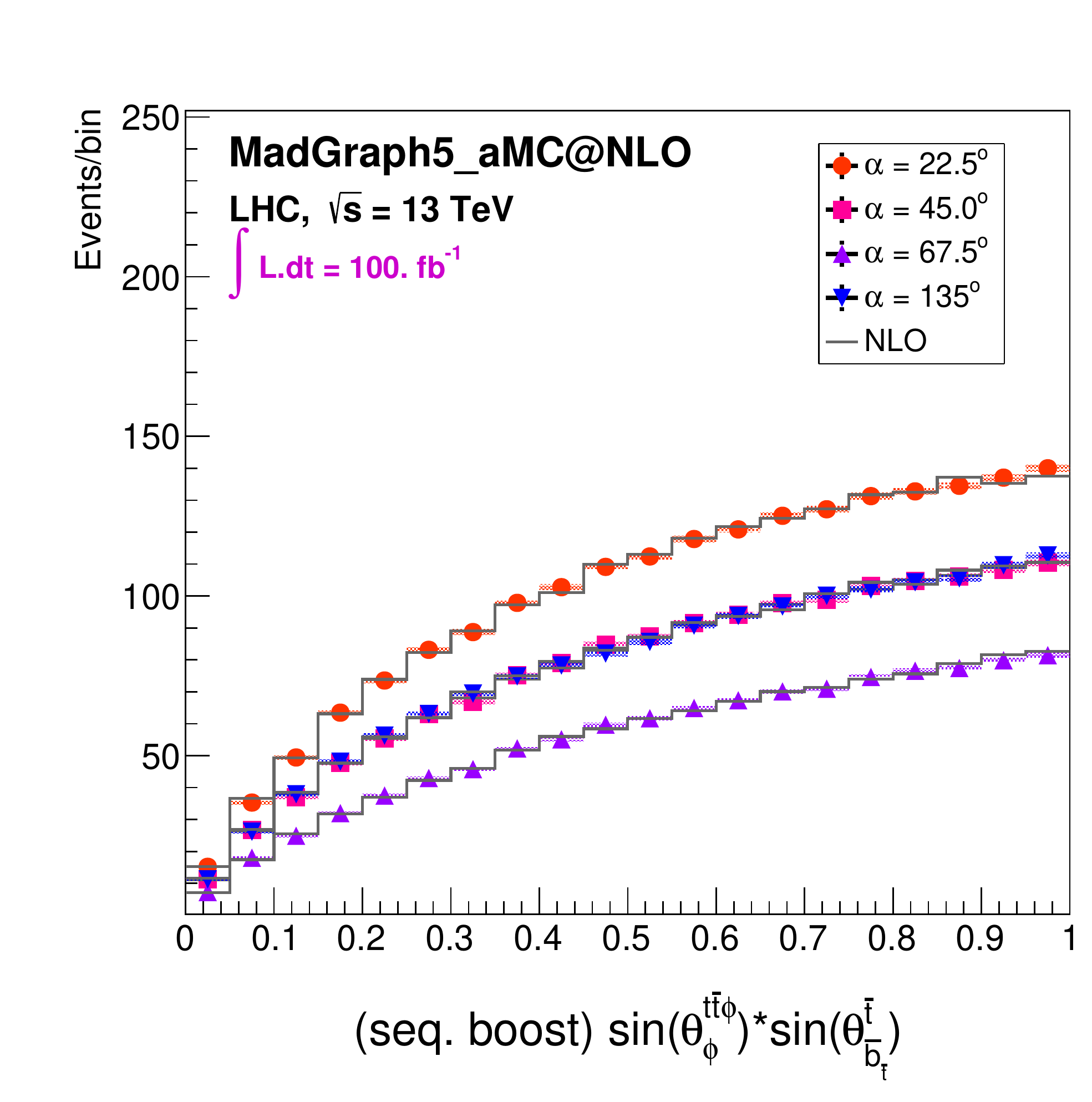}\\
		\caption{The CP observables reconstructed with the pure CP-even and pure CP-odd signals together with the interference terms at parton level with shower effects (with no cuts applied), for different mixing angles ($\alpha = 22.5^{o},45.0^{o},67.5^{o}$ and $135^{o}$), are shown against the corresponding NLO generations. Good agreement between the observables when reconstructed with the interference terms and the parton level distributions is observed.}
		\label{fig:ttX_rec}
	\end{center}
\end{figure}

The differential production cross section associated to any CP-mixed case of the $t\bar{t}\phi$ signal can be parameterized  as a function of the pure scalar and pure pseudo-scalar differential cross sections, according to

\begin{equation}
\begin{aligned}
d\sigma_{t\bar{t}\phi} = \kappa^2 \, \, d\sigma_\text{CP-even} + \tilde{\kappa}^2 \, \, d\sigma_\text{CP-odd} + \kappa \tilde{\kappa} \, \, d\sigma_\text{int}  \\
\end{aligned}
\label{eq:mixed}
\end{equation}
where $d\sigma_{t\bar{t}\phi}$, $d\sigma_\text{CP-even}$, $d\sigma_\text{CP-odd}$ and  $d\sigma_\text{int}$ correspond to the signal differential cross sections for the CP-mixed, CP-even, CP-odd and interference terms, respectively. 
Although the interference terms do not contribute to the total cross section, they can affect the overall shape of the differential distributions for the different CP-observables. One of the most interesting questions we want to address in this paper is to understand the sensitivity of the different CP-observables to the interference terms in $t\bar{t}\phi$ production at the LHC. By re-arranging the terms in Eq.~\ref{eq:mixed}, it is possible to extract the contribution of the interference terms. This is done by subtracting to the CP-mixed differential cross sections the sum of the CP-even and CP-odd differential cross sections (weighted, respectively, by $\kappa^2$ and $\tilde\kappa^2$) and normalising the differences to $\kappa \times \tilde{\kappa}$ . 
\begin{figure}[H]
	\begin{center}
		\includegraphics[height = 7.0cm]{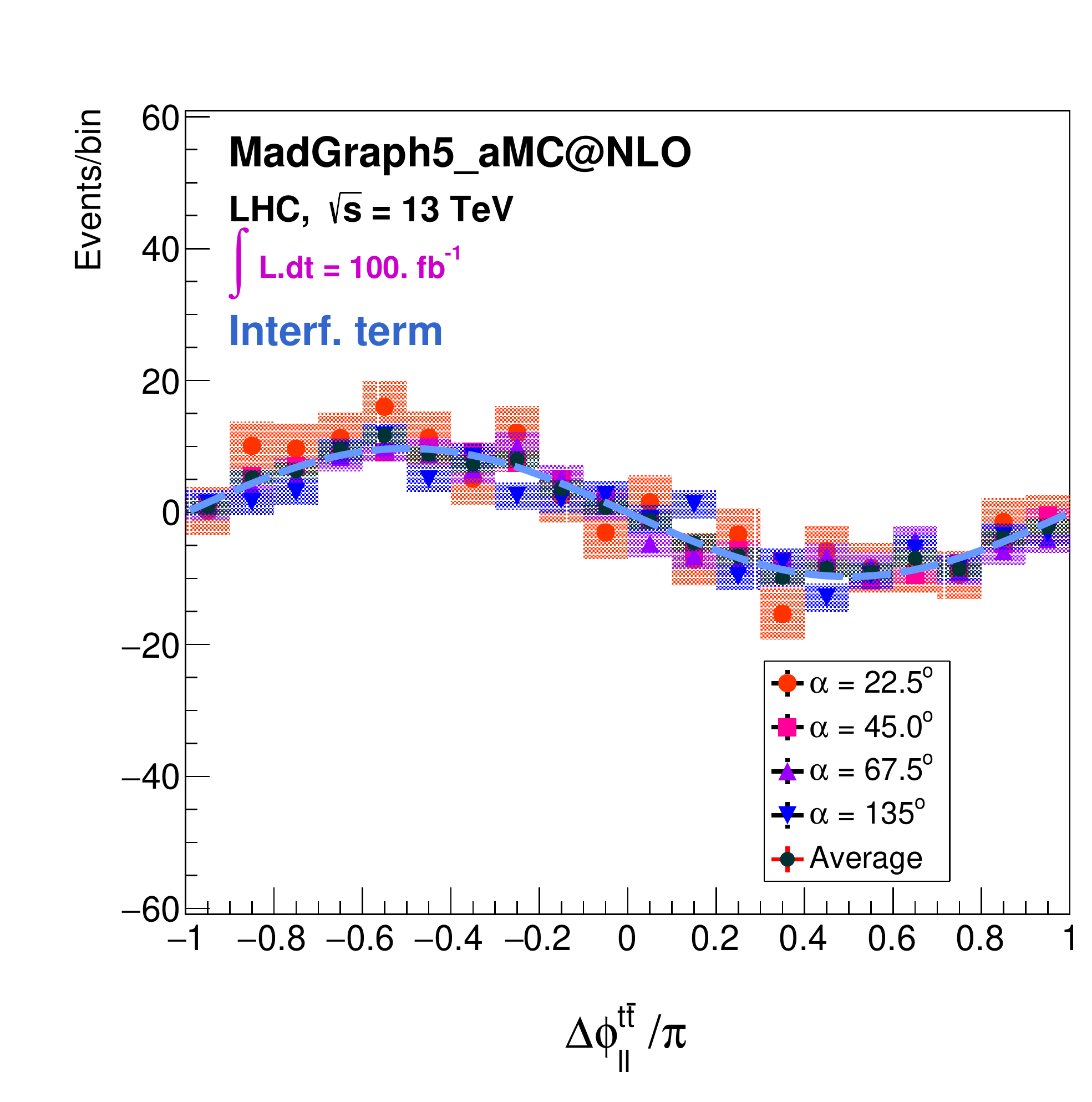}
		\includegraphics[height = 7.0cm]{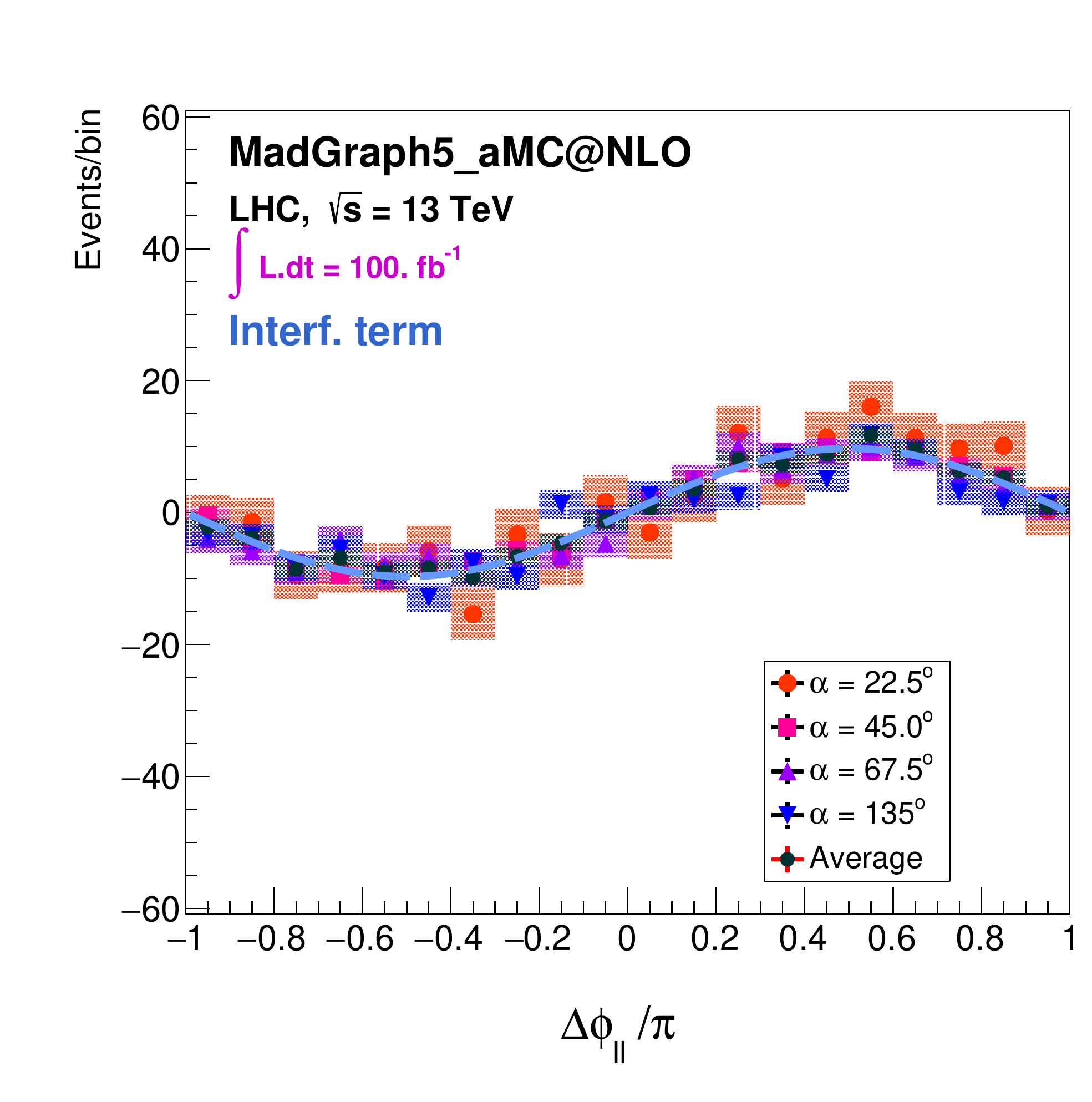} \\
		\includegraphics[height = 7.0cm]{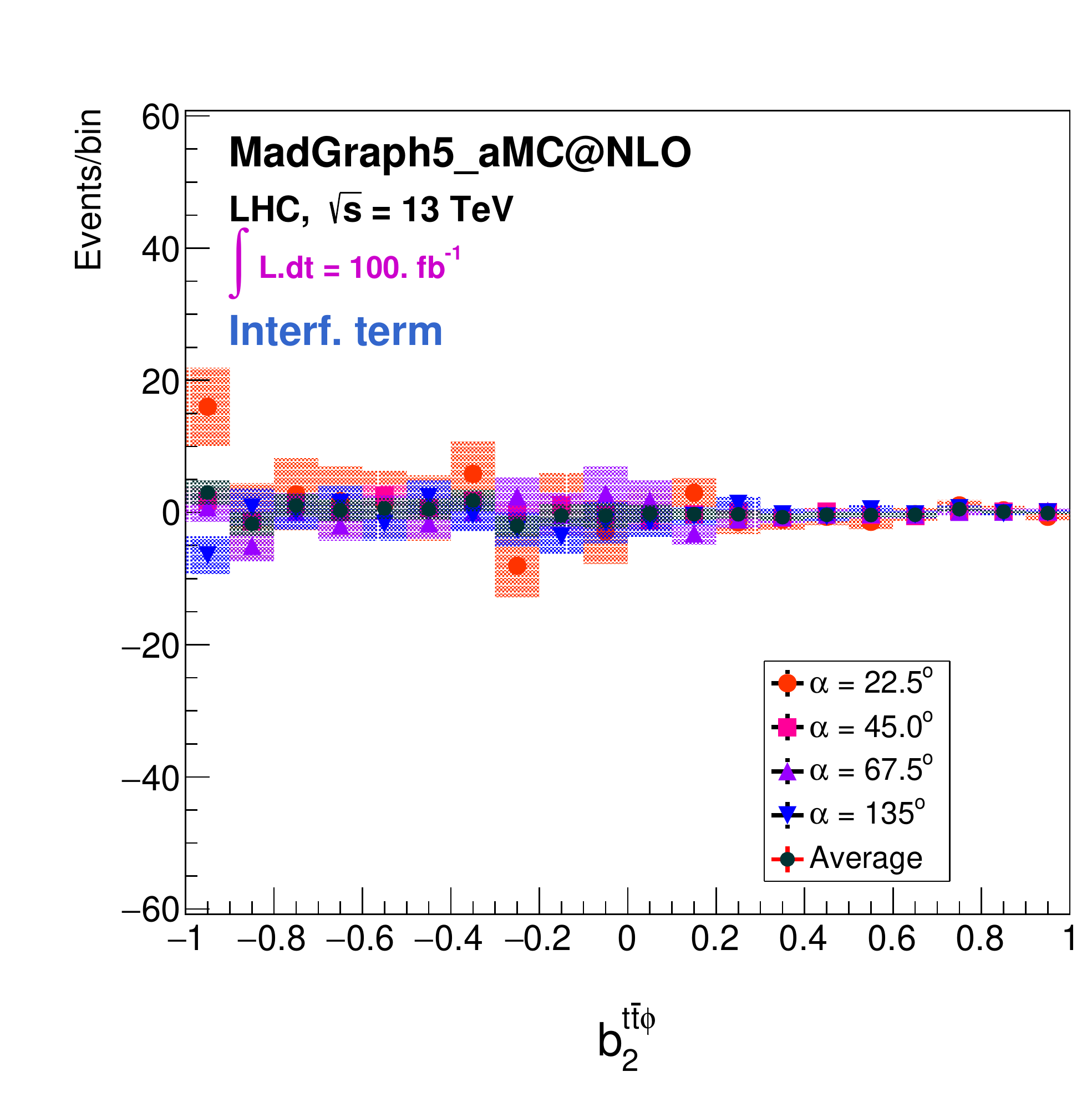}
		\includegraphics[height = 7.0cm]{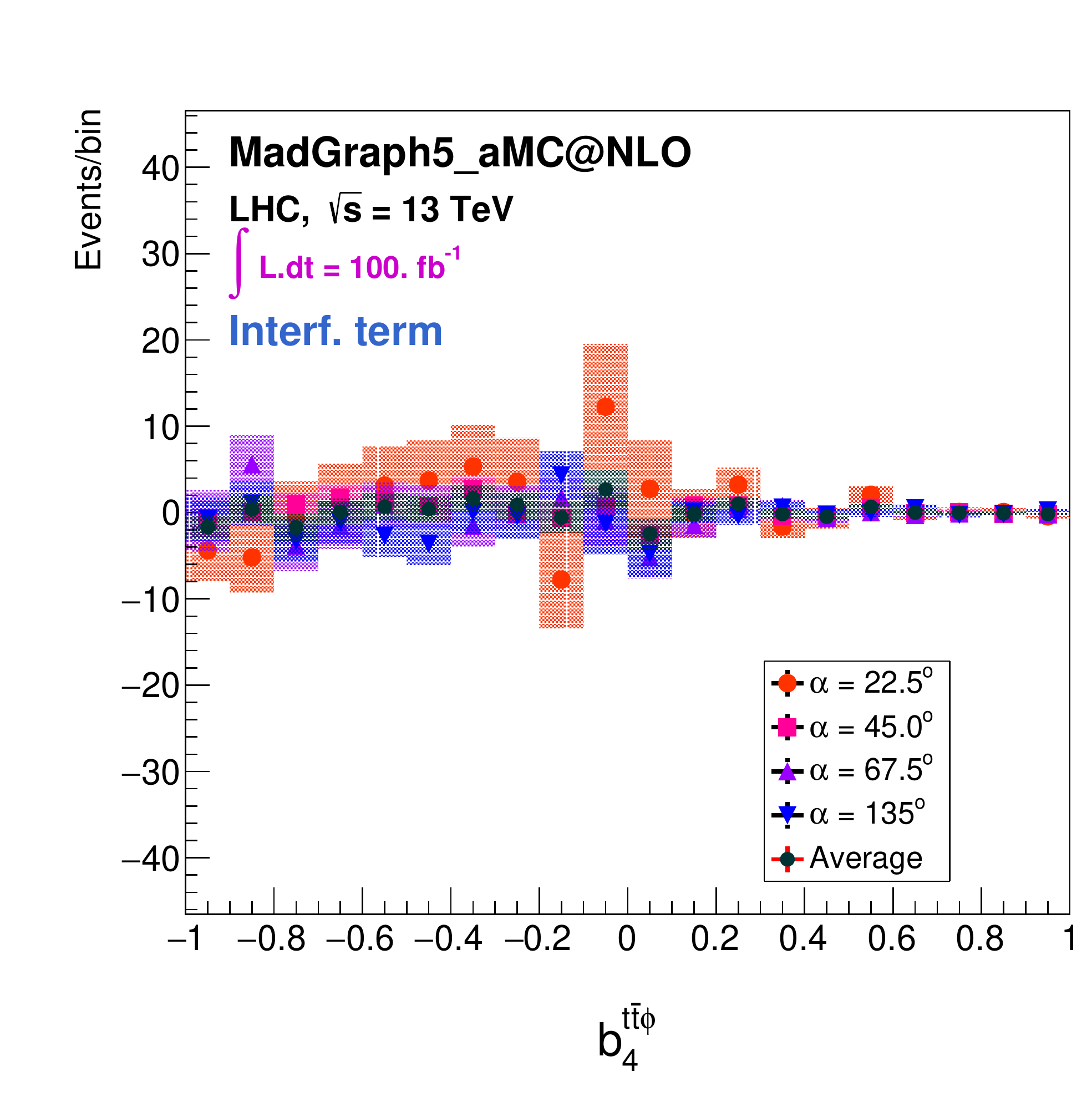} \\
		\includegraphics[height = 7.0cm]{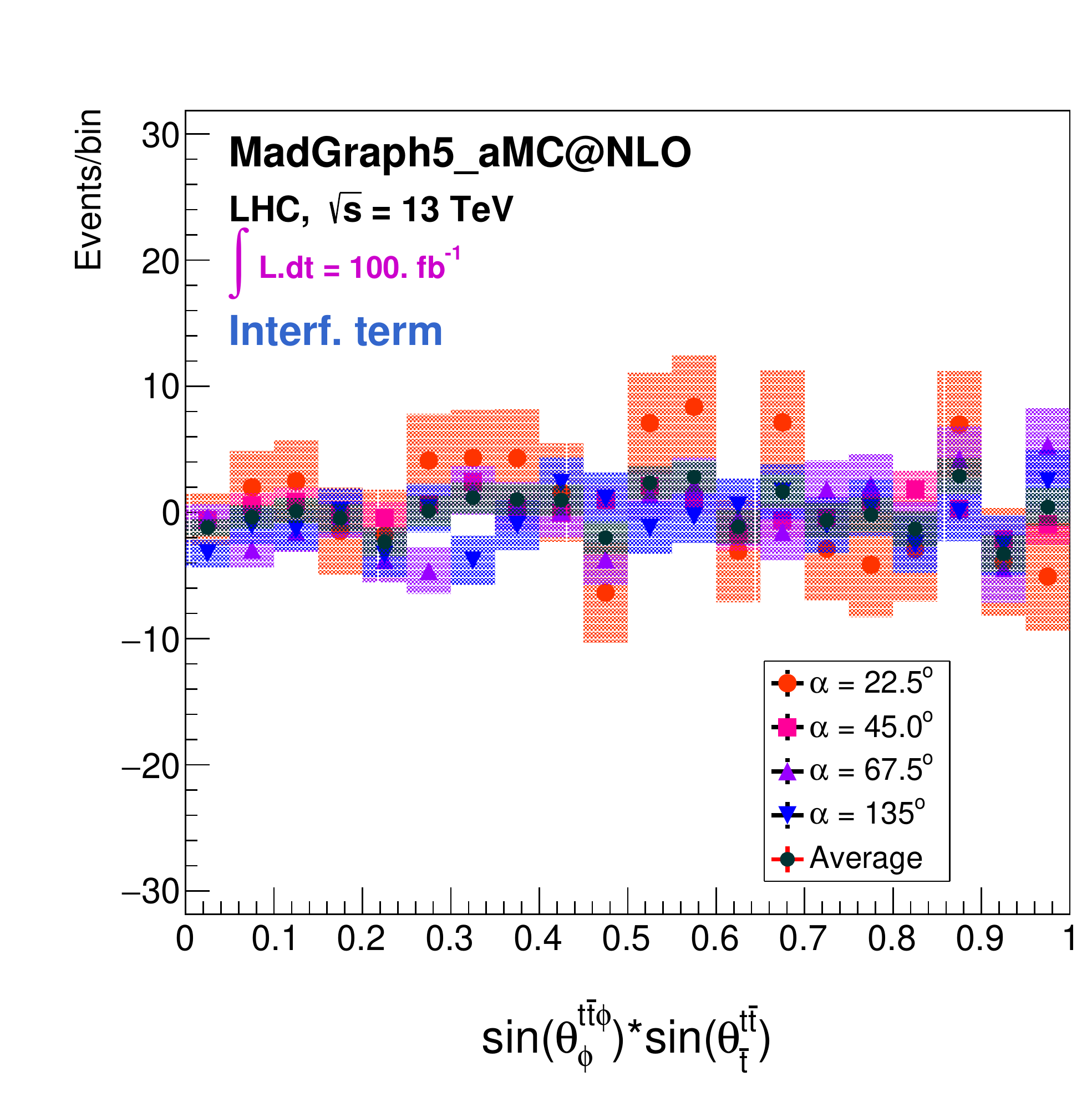}
		\includegraphics[height = 7.0cm]{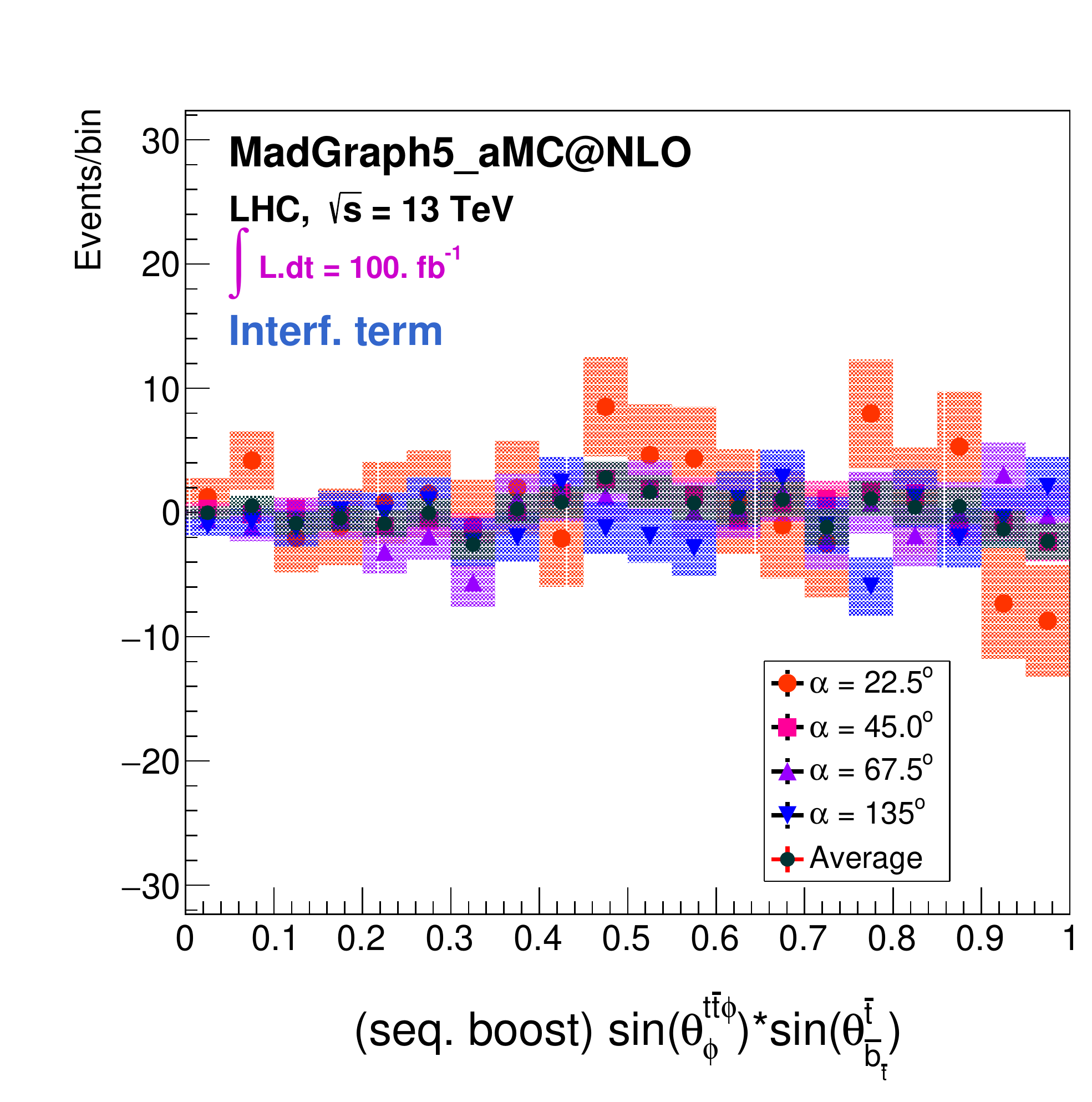}\\
		\caption{Interference term at NLO+Shower effects for our choice of the best CP-sensitive variables in $t\bar{t}\phi$ production at the LHC, for a reference luminosity of 100~fb$^{-1}$. Different $t\bar{t}\phi$ signals, with mixing angles set to $\alpha = 22.5^{o},45.0^{o},67.5^{o}$ and $135^{o}$ are used to extract the interference term.}
		\label{fig:int_parton}
	\end{center}
\end{figure} 
In Fig.~\ref{fig:int_parton} we present the the interference terms at NLO+Shower effects, for the most sensitive CP-observables (as previously discussed)  in $t\bar{t}\phi$ production at the LHC, for a reference luminosity of 100~fb$^{-1}$. The different $t\bar{t}\phi$ signals, with mixing angles set to $\alpha = 22.5^{o},45.0^{o},67.5^{o}$ and $135^{o}$, are used to extract the interference terms. 
From all CP-observables studied, the only ones that show any 
shape dependence are the ones corresponding to $\Delta \phi^{t\bar{t}}_{ll}$ and $\Delta \phi_{ll}$, defined previously. In order to reduce the uncertainty on the interference terms, for each CP-observable, all generated CP-mixed signals are considered to compute an interference term, and their average (also represented in Fig.~\ref{fig:int_parton}) is taken as the interference term for that particular CP-observable. To show that the process of determining the interferences is consistent, a comparison between the CP-mixed signals generated with different mixing angles, and the signals reconstructed using the interference terms, evaluated as described in Eq.~\ref{eq:mixed}, is performed. This comparison is shown in Fig.~\ref{fig:ttX_rec}, where good agreement  is observed not only in what concerns the different differential distributions shapes but also the absolute rates expected for the different CP-mixed signals.

%%%%%%%%%%%%%%%. Event Generation and Kinematic Reconstruction
\section{Event generation, selection and kinematic reconstruction  \label{sec:generation}}
\hspace{\parindent} %forca identacao

\noindent
{\it Monte Carlo Generation and Simulation}\\[3mm]
\noindent
LHC-like signal and background events were generated with \texttt{MadGraph5\_aMC@NLO}~\cite{Alwall:2011uj} for a centre-of-mass energy of 13~TeV. The Higgs Characterization model \texttt{HC\_NLO\_X0}~\cite{Artoisenet:2013puc} was used to generate $pp\to t\bar t\phi$ and single $pp\to t\phi + jets$ signal events. The CP-even ($\phi=H$) and CP-odd ($\phi=A$) signals were generated by setting the mixing angle ($\alpha$) to $\alpha=0^\circ\text{ and }90^\circ$, respectively, following Eq.~\ref{eq:higgscharacter}, with $\kappa_t=1$. 
Additional signal samples with different mixing angles, i.e. $\alpha=$22.5$^\circ$,  45.0$^\circ$, 67.5$^\circ$, 135.0$^\circ$ and 180.0$^\circ$, were also generated. The mass of the scalar was set to $m_\phi=125$~GeV in all signal samples. Only $t\bar{t}\phi$ dileptonic events were considered ($t\bar{t}\rightarrow bW^+\bar{b}W^-\rightarrow b\ell^+\nu_\ell\bar{b}\ell^-\bar{\nu}_{\ell}$) and the scalar boson is set to decay to a pair of $b$-quarks ($\phi\to b \bar b$).
Backgrounds from SM $t\bar{t}$ + 3~jets, $t\bar{t}V$ + jets, single top quark production ($t$-, $s$- and $Wt$-channels), $W$($Z$) + 4~jets, $W$($Z$)$b\bar{b}$ + 2~jets and $WW, ZZ, WZ$ diboson processes were also generated using \texttt{MadGraph5\_aMC@NLO}. 
Following event generation and hadronization by \texttt{PYTHIA}~\cite{Sjostrand:2006za}, all signal and background events were passed through a fast simulation of a typical LHC detector using DELPHES~\cite{deFavereau:2013fsa}. Further details on the event generation and detector simulation can be found in~\cite{Azevedo:2020vfw}. The analysis of signal and background events follows using the \texttt{MadAnalysis5}~\cite{Conte:2012fm} framework.\\

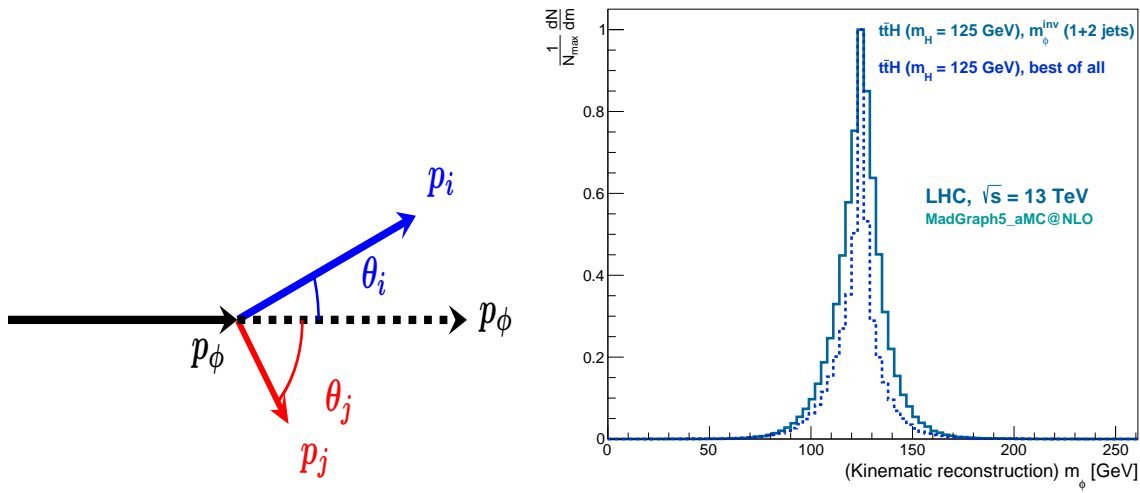
\begin{figure}[H]
	\centering
	\resizebox{7.0cm}{4.5cm}{%
	    \begin{tikzpicture}[scale=1.10]
	    %\draw[line width=2pt,blue,-stealth] (2.83,0)--(4.83,2) node[anchor=south west]{$\boldsymbol{p_1}$};
	    \draw[line width=2pt,blue,-stealth] (2.83,0)--(5.03,0.8) node[anchor=south west]{$\boldsymbol{p_i}$};
	    %\draw[line width=2pt,red,-stealth](2.83,0)--(4.83,-2) node[anchor=north west]{$\boldsymbol{p_2}$};
	    \draw[line width=2pt,red,-stealth](2.83,0)--(3.46,-0.8) node[anchor=north west]{$\boldsymbol{p_j}$};
	    \draw[line width=2pt,black,-stealth](0,0)--(2.83,0) node[anchor=north east]{$\boldsymbol{p_\phi}$};
	    \draw[line width=2pt,black,-stealth, dashed](2.83,0)--(5.66,0) node[anchor=west]{$\boldsymbol{p_\phi}$};
	    %\draw[line width=1pt,blue] (3.83,0) arc (0:45:1) node[right=5mm]{$\boldsymbol{\theta_1}$};
	    \draw[line width=1pt,blue] (3.83,0) arc (0:19.98:1) node[right=5mm]{$\boldsymbol{\theta_i}$};
	    %\draw[line width=1pt,red] (3.63,0) arc (0:-45:0.8) 	
	    \draw[line width=1pt,red] (3.63,0) arc (0:-51.78:0.8) node[right=5mm]{$\boldsymbol{\theta_j}$};
	    \end{tikzpicture}
	}
	\hspace*{0mm}\includegraphics[height = 7.1cm]{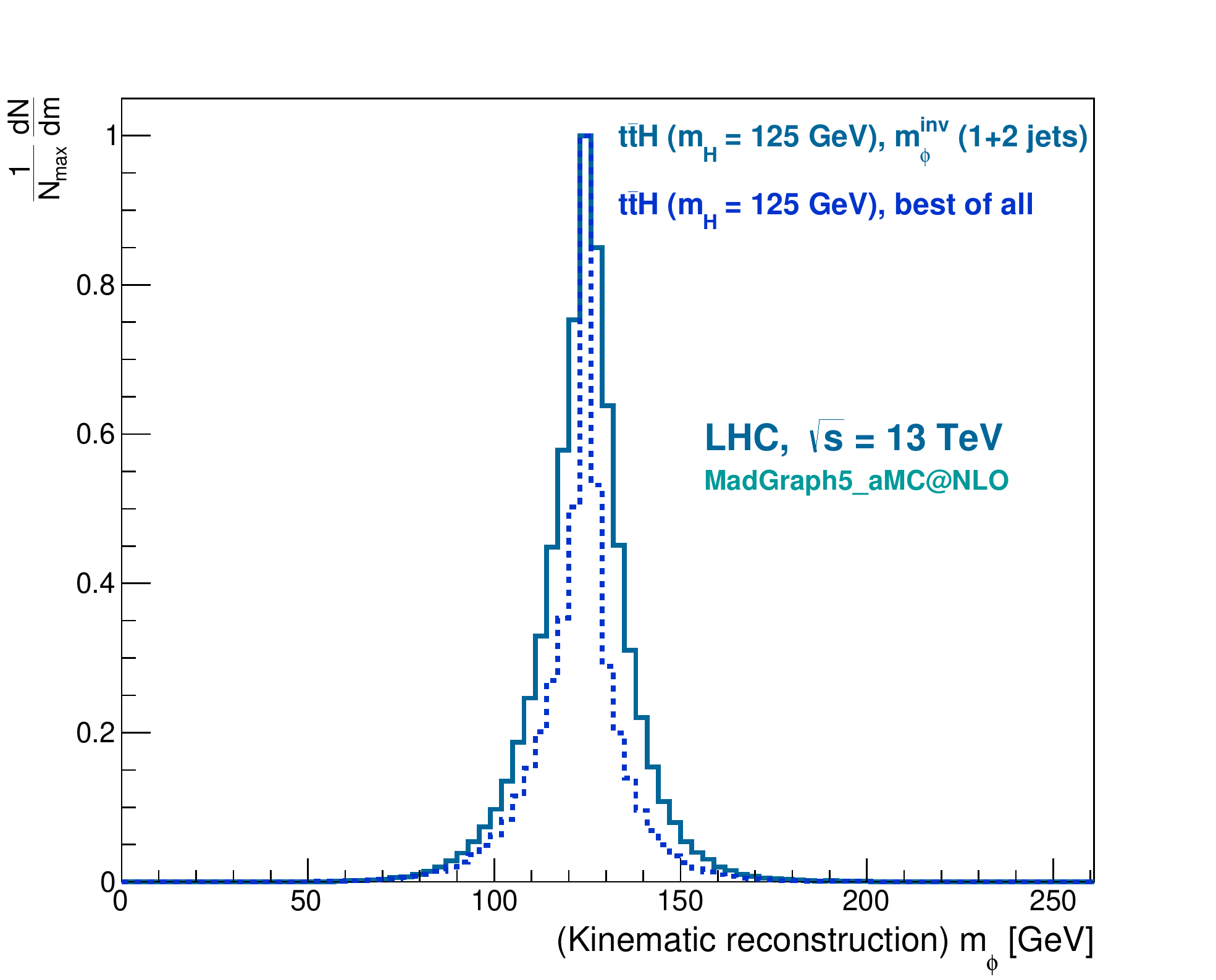}
	\caption{ (left) Schematic representation of the $\phi$ boson decay and angles between the Higgs and its decay products; (right) SM Higgs boson mass distribution after kinematic reconstruction, for $m_\phi= 125$~GeV. The solid line shows the best invariant Higgs mass from one or two jets i.e., $m^{inv}_\phi$ (1+2 jets) = $m^{inv}_\phi$ (1 jet) or $m^{inv}_\phi$ (2 jets), and the dashed line represents the best of all methods (best of all) introduced in~\cite{Azevedo:2020fdl}, $\phi=H$.}
	\label{fig:scheme}
\end{figure}

\noindent
{\it Selection and Kinematic Reconstruction}\\[3mm]
\noindent
Events are selected by requiring at least four jets and two opposite charge leptons in the final state, with pseudo-rapidities ($\eta$) below 2.5 and transverse momenta ($p_T$) above 20~GeV. As was done in~\cite{Azevedo:2020fdl}, we start the kinematic reconstruction by evaluating the mass of the $\phi$ boson and the jets associated to it. We use the usual invariant mass of pairs of jets $(1,2)$ to evaluate the scalar boson mass,  $m^{inv}_\phi=\sqrt{(p_1+p_2)^2}$~\footnote{$p_1$ and $p_2$ are the 4-momentum of jets 1 and 2, respectively.}, as well as a new method that uses the jets polar angles, 
\begin{equation}
\begin{aligned}
m^{(i)}_\phi = |\vec{p}_1| \, \sqrt{2 \, \frac{\sin \theta_1}{\sin \theta_2} \bigg[1 - \cos (\theta_1 + \theta_2)\bigg]}, \, \,  \\ 
\end{aligned}
\label{eq:mass_new}
\end{equation}
where $\vec{p}_1$ corresponds to the 3-momentum of jet $1$ and $\theta_1$ ($\theta_2$) is the polar angle of jet 1 (jet 2), with respect to the direction of flight of the two jet system, assumed to be the scalar boson direction. 
Eq.~\ref{eq:mass_new} reflects the momentum conservation of a scalar boson decaying to two jets as represented in Fig.~\ref{fig:scheme} (left). This method allows the mass measurement to be performed with only the knowledge of the 3-momentum of one jet, together with two polar angles from both jets. As angles tend to be well resolved by the LHC experiments, this method avoids momentum resolution effects from two jets to spoil the mass measurement (as usually happens in an invariant mass calculation), with only one jet momentum being necessary.

For completeness, a one-jet invariant mass reconstruction ($m_\phi^{inv}$) is also implemented as a possible outcome for the Higgs boson decay, in particular when the decay jets overlap, but this is largely irrelevant in our analysis. 
From all methods used to evaluate the mass, the one giving the closest result to the Higgs boson mass ($m_\phi = 125$~GeV) is chosen. The jets used by this method are associated, by the kinematic reconstruction, to the scalar boson decay partons. In Fig.~\ref{fig:scheme} (right), we can see that the mass resolution, due to the impact of the new method, improves by roughly a factor two compared with the usual invariant mass measurement. The correct identification of the remaining jets as coming from the top quarks decays, and the reconstruction of the $t\bar{t}$ system (which includes the neutrinos, the $W^{\pm}$ bosons and the $t$ and $\bar{t}$ quarks), is performed by a kinematic fit detailed in~\cite{Azevedo:2020fdl}. 

Events are further selected by accepting the ones with two isolated leptons with opposite charges and invariant mass, $m_{\ell^+ \ell^-}$, outside 
a window of 10~GeV around the $Z$ boson mass ($m_Z=91$~GeV), to avoid contamination from the $Z$ + jets background. Only events with at least 3 jets tagged as coming from the hadronization of $b$-quarks (labeled as {\it $b$-tagged}), are accepted. Further details on efficiencies, total number of events and kinematic distributions, can be found in~\cite{Azevedo:2020fdl}.

In Tab.~\ref{tab:AsymExp},  we present the asymmetries for the $t\bar{t}\phi$ signal as a function of the  mixing angle $\alpha$, as well as for the dominant background $t\bar{t}b\bar{b}$ after event selection and kinematic reconstruction.
The variable $x_c$ represent the point chosen for the calculation of the asymmetry while $A_c$ is the asymmetry relative to that point. We can see that there are significant differences between the asymmetries for the pure scalar ($\alpha=0.0^\circ$) and for pseudo-scalar ($\alpha=90.0^\circ$) scenarios for several asymmetries. As we will show later, an appropriate choice will maximise the difference between the asymmetries from the CP-even and CP-odd pure signal cases.

\begin{table}[h]
\renewcommand{\arraystretch}{1.3}
\begin{center}
 \hspace*{-2mm}
  \begin{tabular}{c|c|ccccccc|c}
    \toprule
                &         & \multicolumn{8}{c}{ MadGraph5 @ NLO+Shower (after selection and rec.)}        \\[-1mm]
    Asymmetries &  $x_c$  & \multicolumn{7}{c}{ $t\bar{t}\phi$ signal mixing angle $\alpha$ (deg.)} &    \\
                &         &  $0.0^\circ$ & $22.5^\circ$ & $45.0^\circ$ & $67.5^\circ$ & $90.0^\circ$ & $135.0^\circ$ & $180.0^\circ$ &  $t\bar{t}b\bar{b}$ \\
    \midrule 
% NLO+Shower values 
      \hspace*{-3mm}$A_c[b_2^{t\bar{t}\phi}]$  
        & $-0.30$ & $-0.12$ & $-0.11$ & $-0.01$ & $+0.16$ & $+0.24$ & $-0.01$ & $-0.13$	         & $-0.03$   \\
      \hspace*{-3mm}$A_c[b_4^{t\bar{t}\phi}]$ 
        & $-0.50$ & $+0.30$ & $+0.28$ & $+0.18$ & $+0.03$ & $-0.06$ & $+0.17$ & $+0.32$	 & $+0.26$   \\
      \hspace*{-3mm}$A_c[\sin (\theta^{t\bar{t}\phi}_\phi)*\sin(\theta^{t\bar{t}}_{\bar{t}})]$ 	
        & $+0.70$ & $-0.26$ & $-0.26$ & $-0.24$ & $-0.22$ & $-0.19$ & $-0.25$ & $-0.25$	          & $-0.37$   \\
      \hspace*{-3mm}$A_c[\sin (\theta^{t\bar{t}\phi}_\phi)*\sin(\theta^{\bar{t}}_{\bar{b}_{\bar{t}}})$]	
        & $+0.60$ & $+0.01$ & $-0.01$ & $-0.02$ & $-0.02$ & $-0.01$ & $-0.03$ & $-0.03$	 & $-0.22$   \\[-2mm]
      \hspace*{-3mm}\text{ (seq. boost)}	
        &  &  &  &  &  &  &  & 	 &   \\
    \bottomrule
  \end{tabular}
\caption{Asymmetries for the $t\bar{t}\phi$ signal as a function of the  mixing angle $\alpha$, as well as for the dominant background $t\bar{t}b\bar{b}$ after event selection and kinematic reconstruction, 
are shown for several observables. The variable $x_c$ represent the point chosen for the calculation of the asymmetry while $A_c$ is the asymmetry relative to that point. Significant differences between the asymmetries for the pure scalar ($\alpha=0.0^\circ$) and pseudo-scalar ($\alpha=90.0^\circ$) cases are observed for several asymmetries.}
\label{tab:AsymExp}
\end{center}
\end{table}

\begin{figure}[H]
	\begin{center}
		\includegraphics[height = 7.0cm]{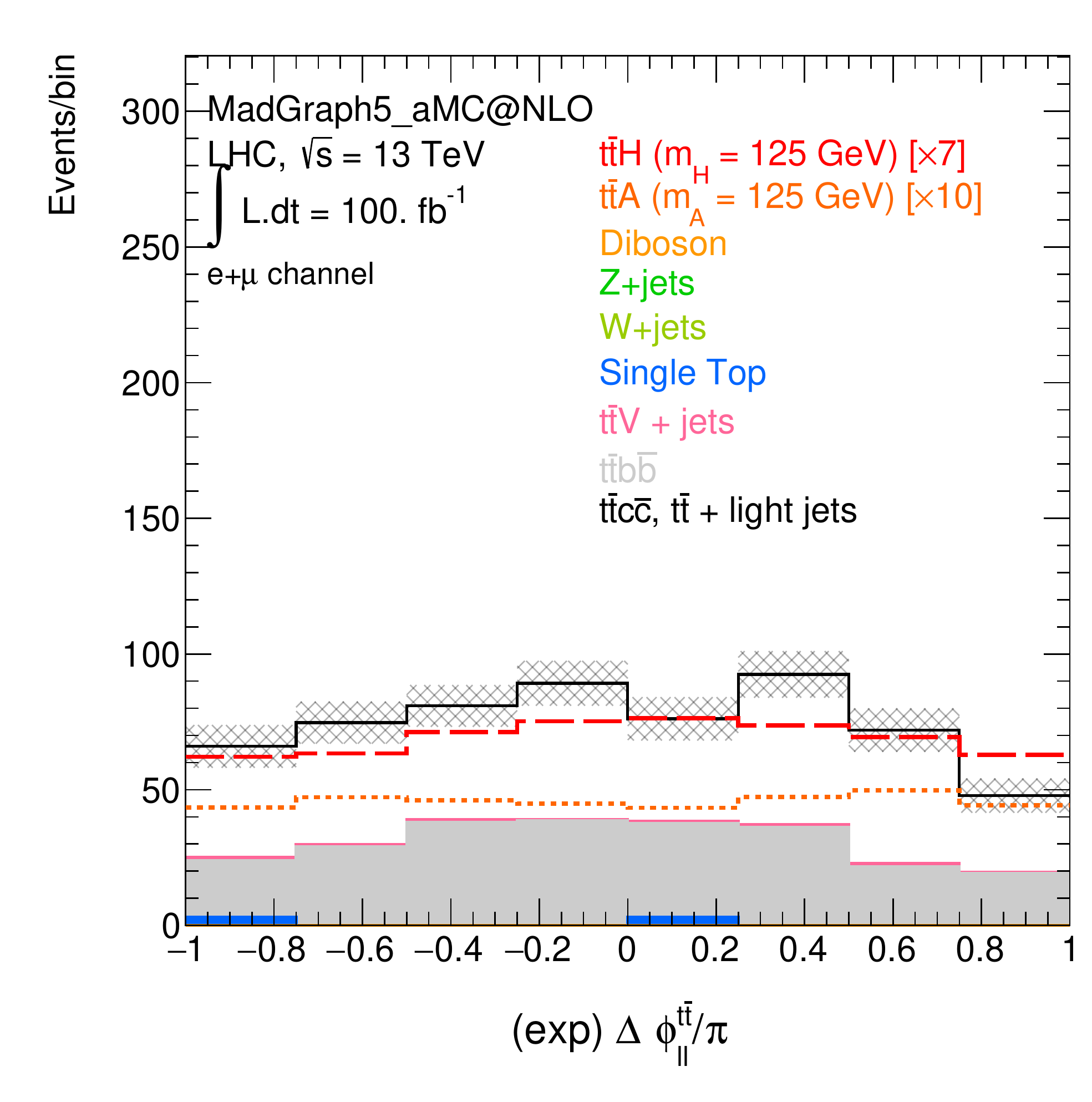}	
		\includegraphics[height = 7.0cm]{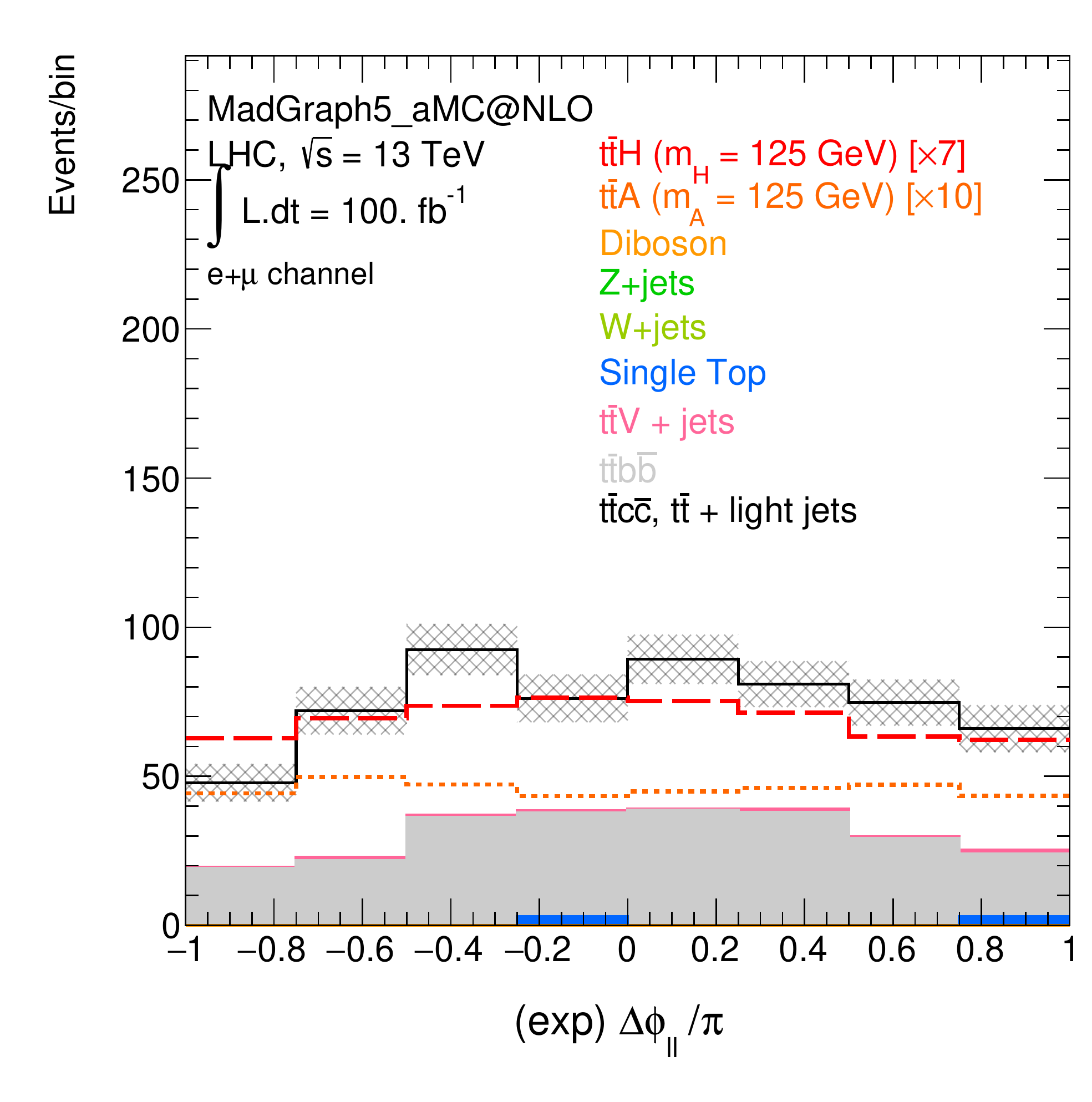}		\\
		\includegraphics[height = 7.0cm]{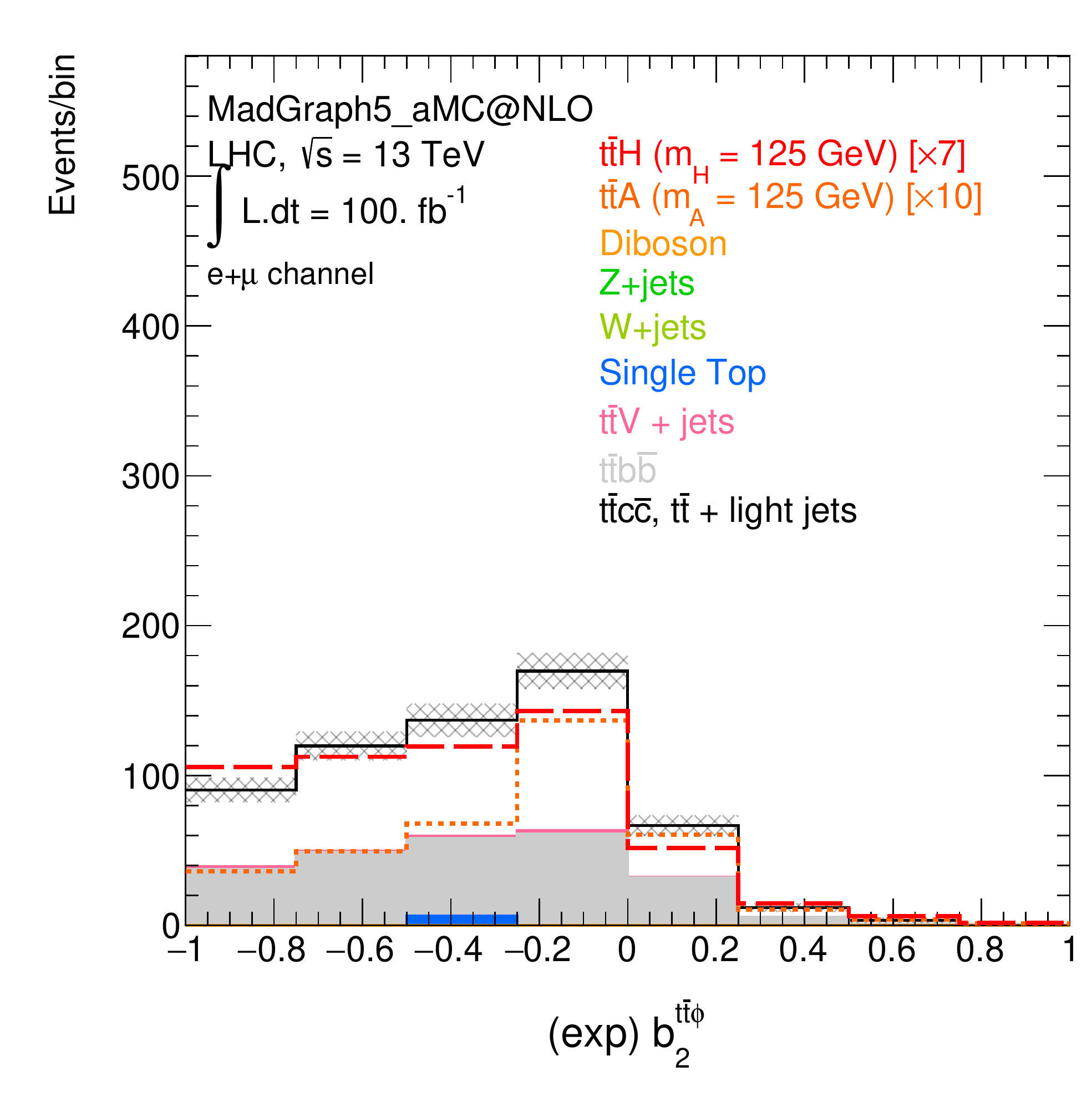}		
		\includegraphics[height = 7.0cm]{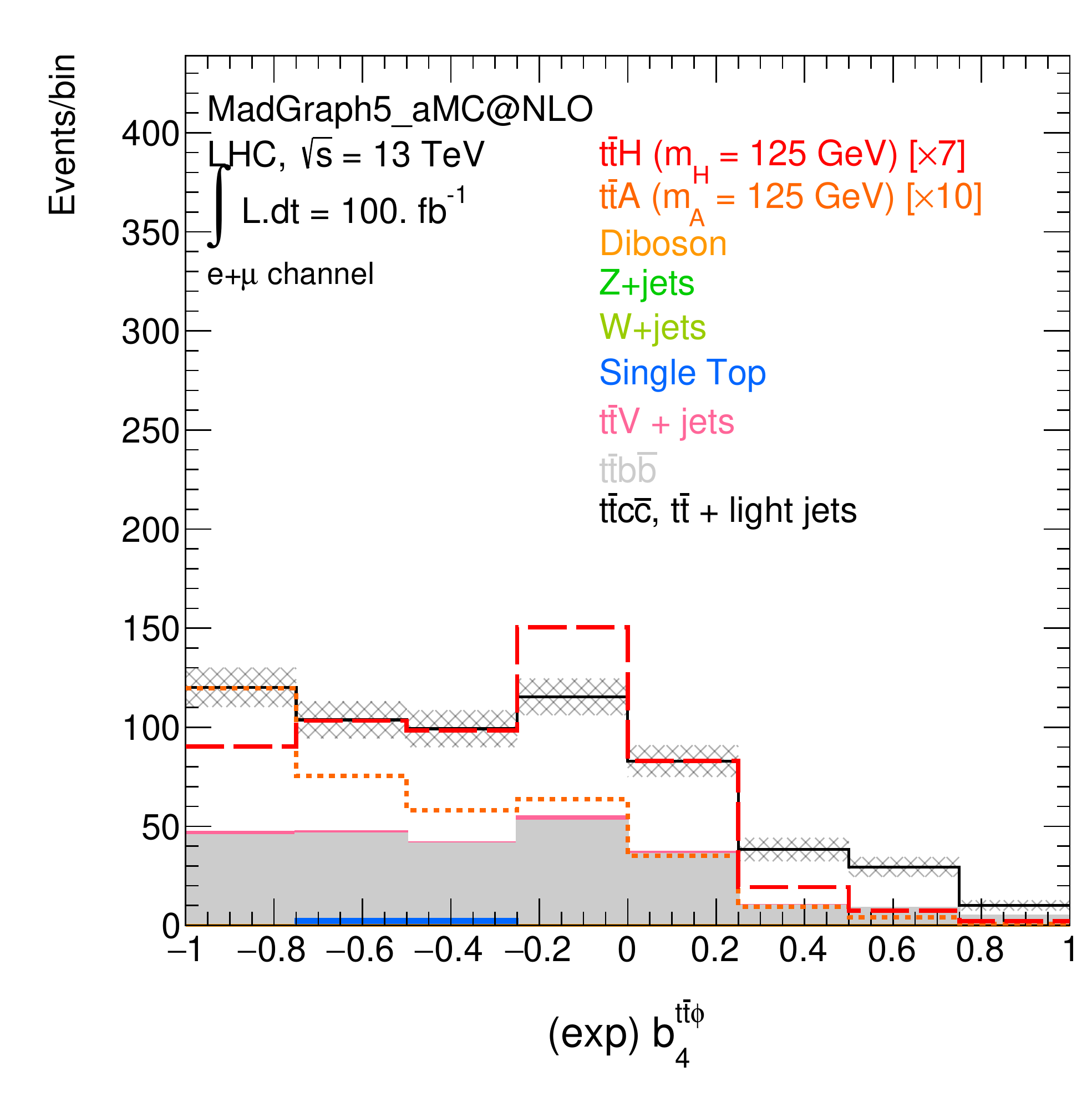}		\\
		\includegraphics[height = 7.0cm]{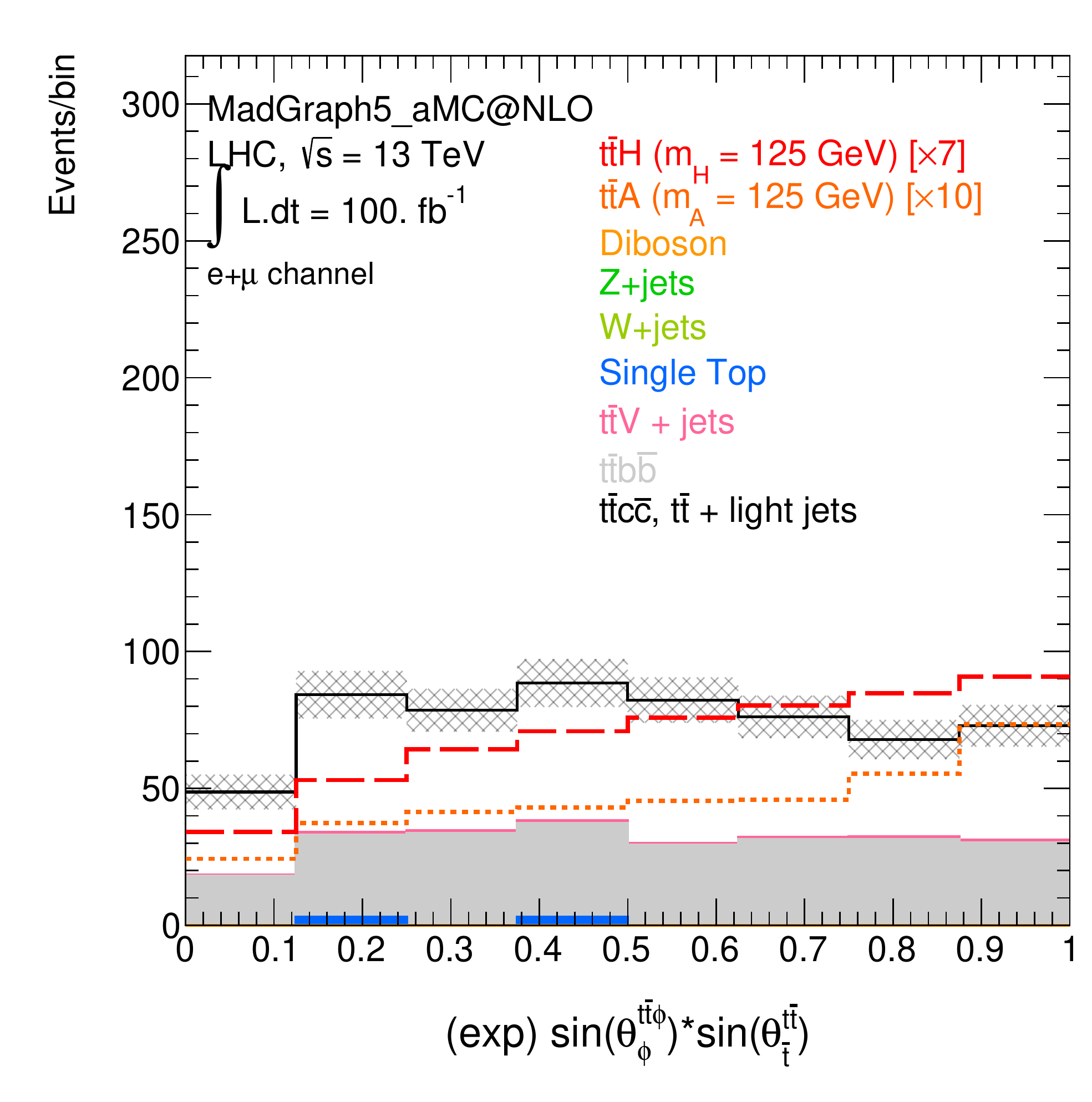}
		\includegraphics[height = 7.0cm]{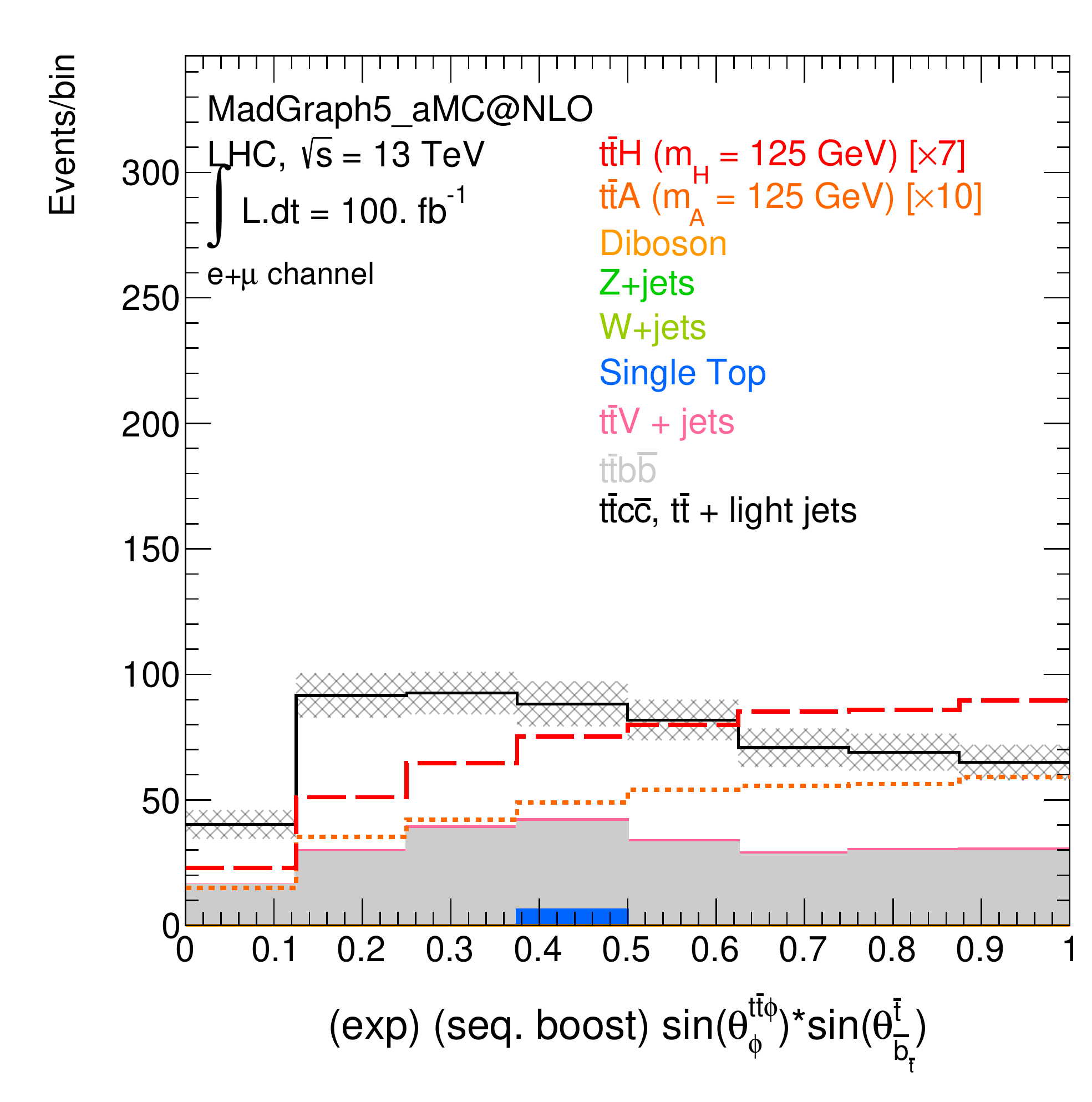}\\
		\caption{Angular distributions are represented, after event selection and kinematic reconstruction, for several sensitive CP variables, in $t\bar{t}\phi$ production at the LHC, for a reference luminosity of 100~fb$^{-1}$. The $t\bar{t}\phi$ CP-even and CP-odd signals, are scaled by a factor of 7 and 10, respectively, for convenience.}
		\label{fig:angular_exp}
	\end{center}
\end{figure}

\begin{figure}[H]
	\begin{center}
		\includegraphics[height = 7.0cm]{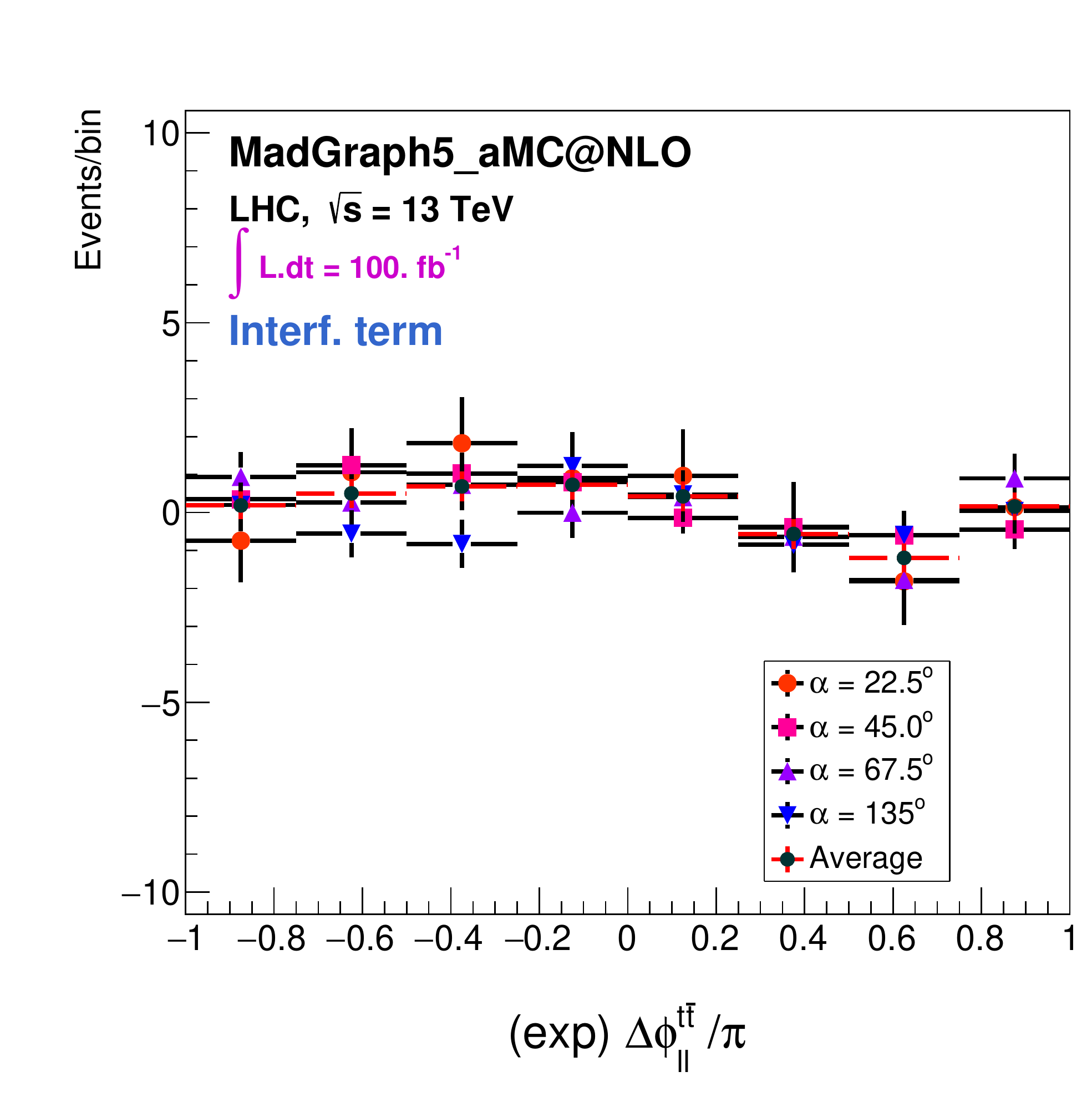}	
		\includegraphics[height = 7.0cm]{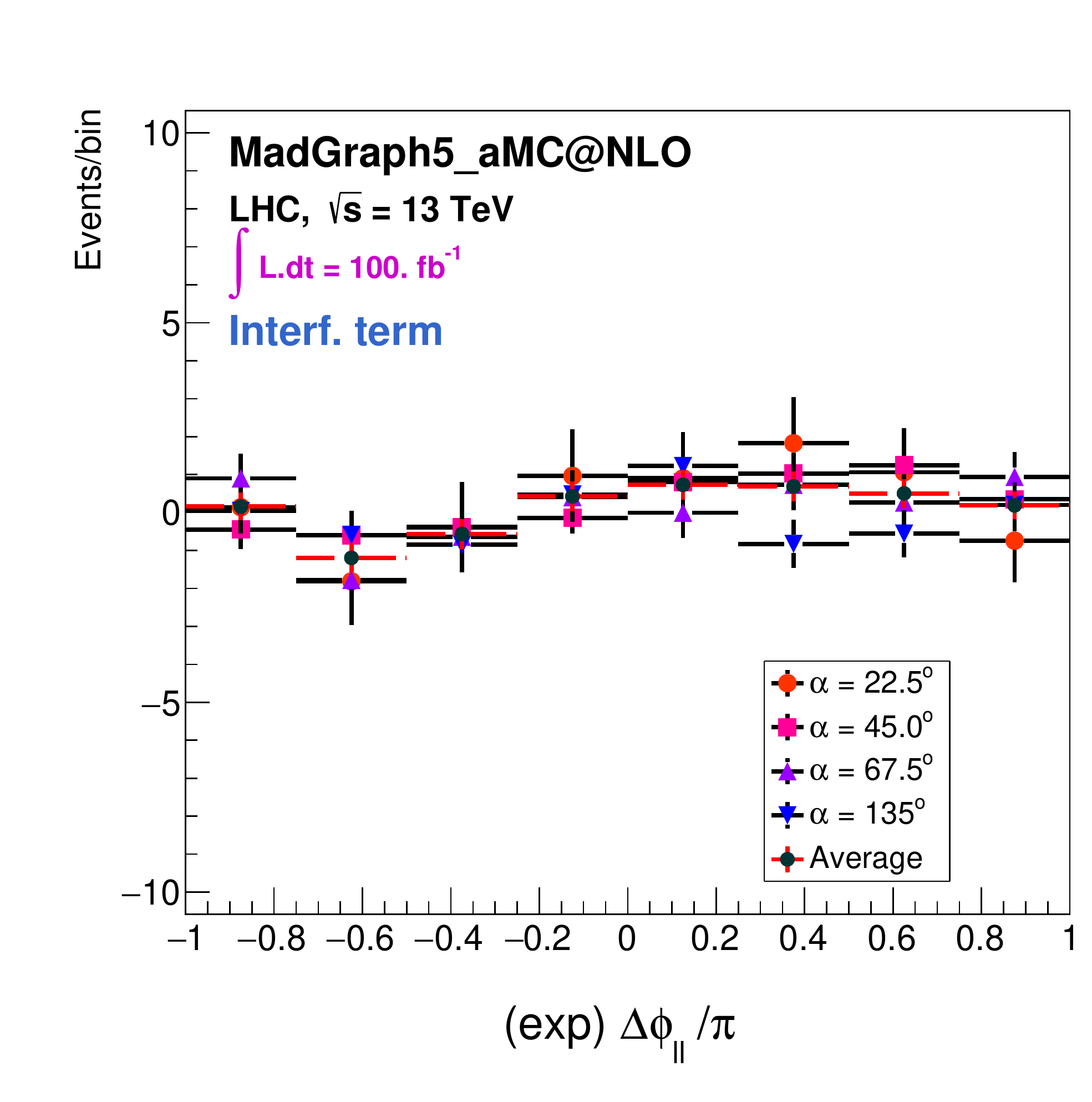}		\\
		\includegraphics[height = 7.0cm]{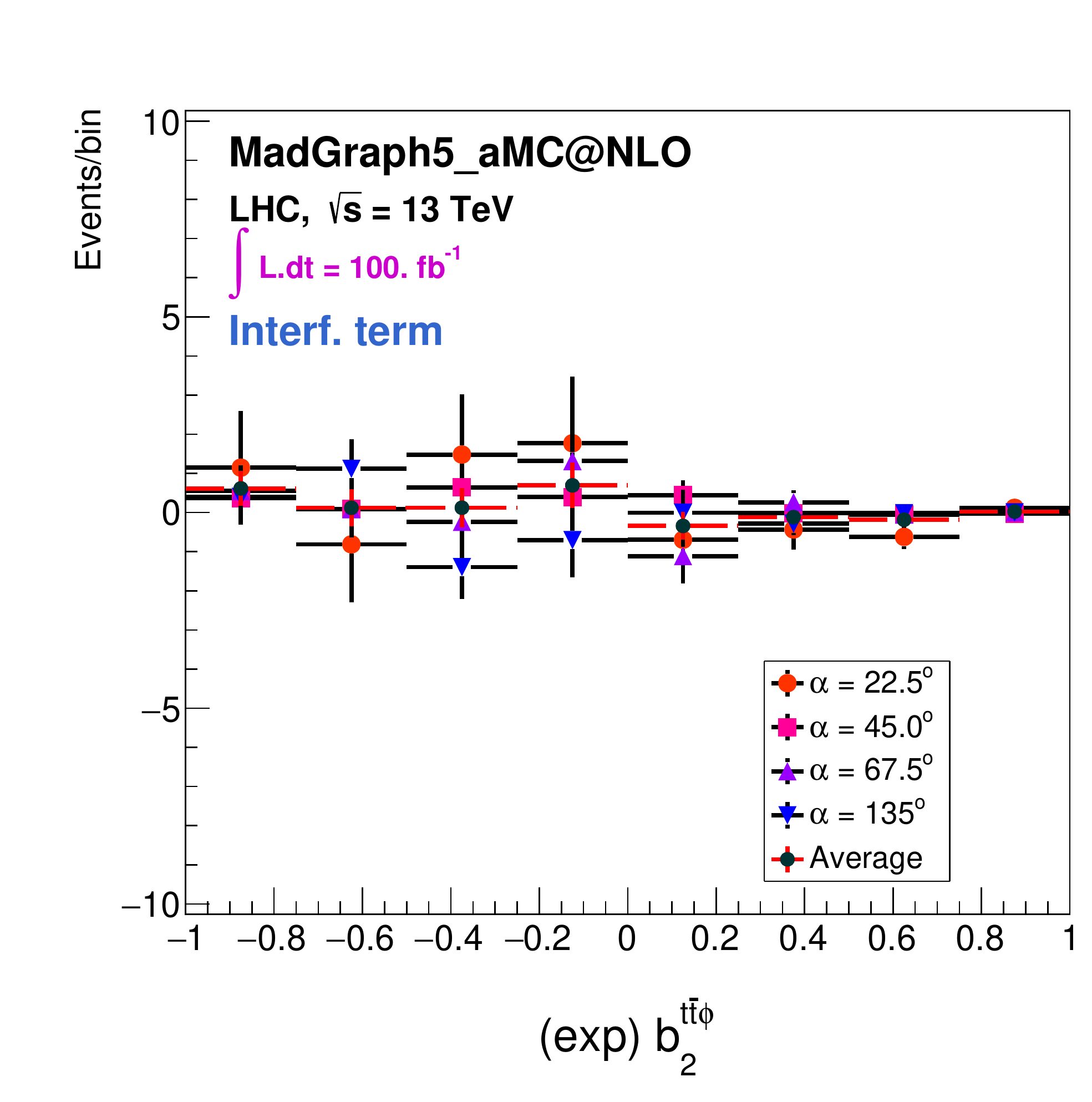}		
		\includegraphics[height = 7.0cm]{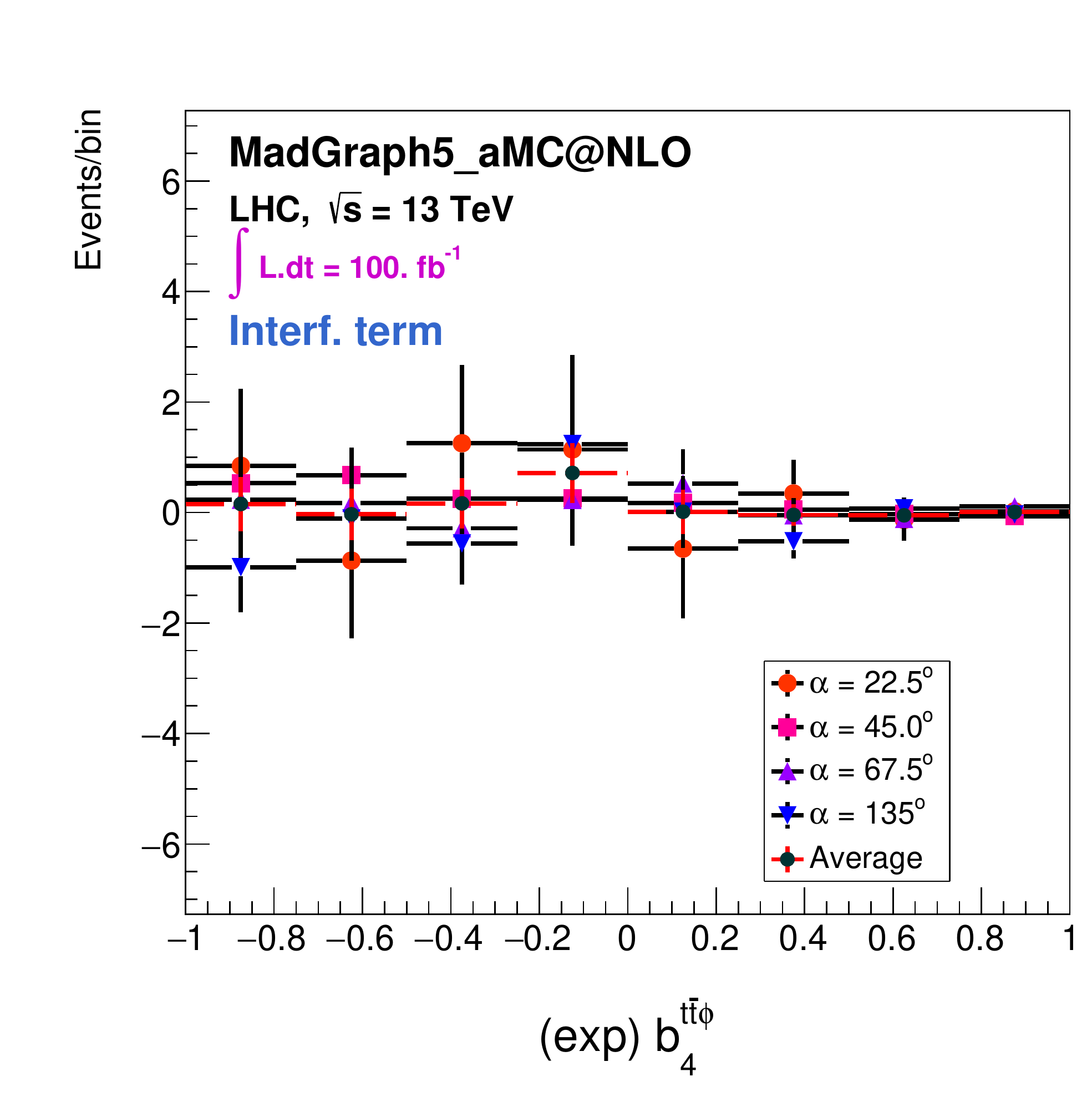}		\\
		\includegraphics[height = 7.0cm]{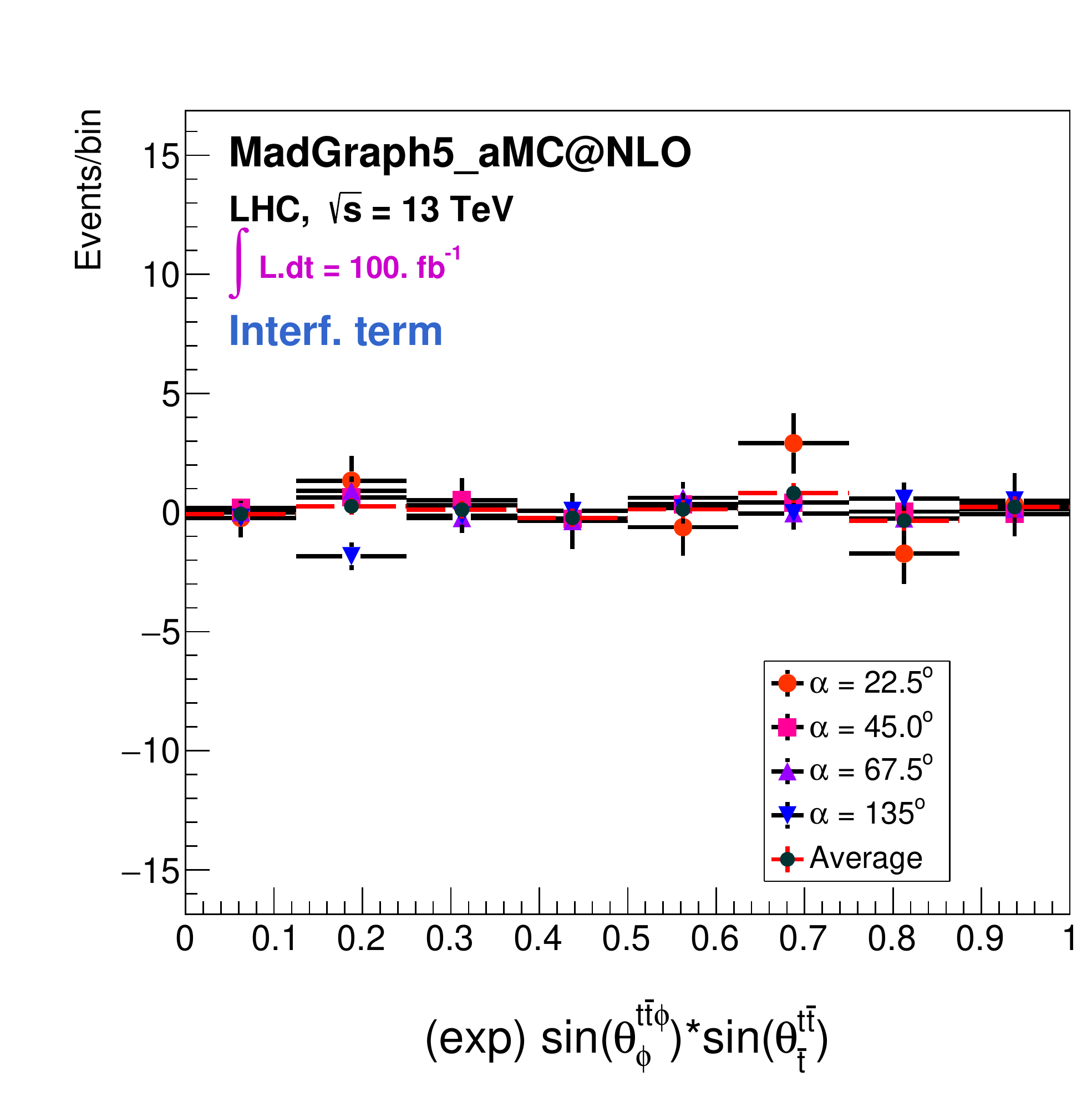}
		\includegraphics[height = 7.0cm]{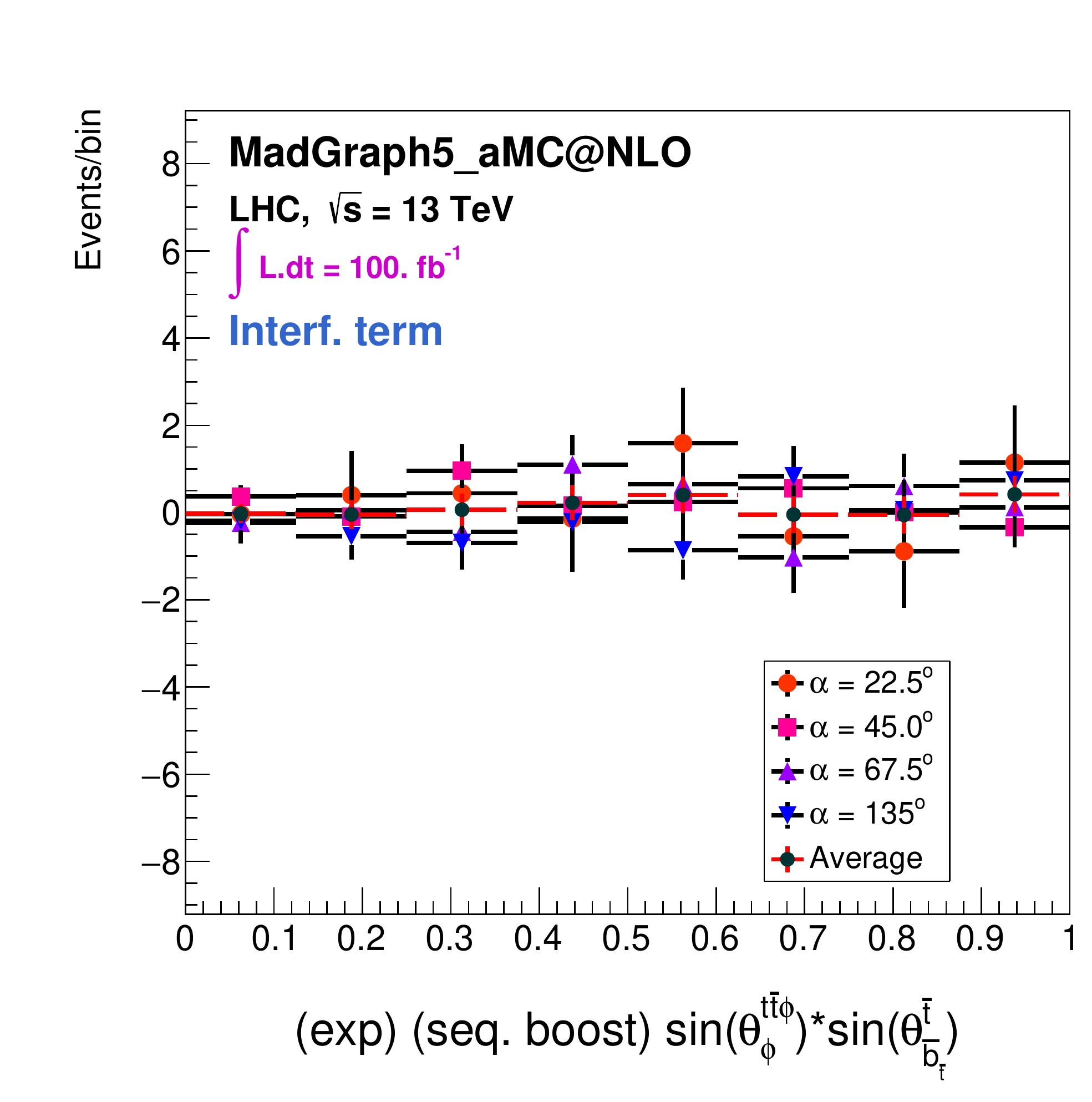}\\
		\caption{The scalar-pseudoscalar interference term is represented, after event selection and kinematic reconstruction, for several sensitive CP variables, in $t\bar{t}\phi$ production at the LHC, for a reference luminosity of 100~fb$^{-1}$. Different $t\bar{t}\phi$ signals, with mixing angles set to $\alpha = 22.5^{o},45.0^{o},67.5^{o}$ and $135^{o}$ are used to extract the interference term.}
		\label{fig:int_exp}
	\end{center}
\end{figure}

In Fig.~\ref{fig:angular_exp}, the differential distributions of the CP-observables are presented, for a reference luminosity of 100~fb$^{-1}$ at the LHC. The $t\bar{t}\phi$ CP-even and CP-odd signals are scaled by a factor of 7 and 10, respectively, for displaying convenience. Although detector effects are clearly visible as distortions in the distributions shapes, the overall trend of the angular distribution is largely preserved. 
On the contrary, for the interference terms, evaluated in the same way as previously at parton level, few events are observed in the differential distributions, see Fig.~\ref{fig:int_exp}. Although some shapes are still noticeable in the $\Delta\phi_{ll}^{t\bar{t}}$ and $\Delta\phi_{ll}$ angular distributions, given the low number of events in every single bin, it is unlikely that the dileptonic $t\bar{t}\phi$ analysis will ever be sensitive to this term. Looking for example into $\Delta\phi_{ll}^{t\bar{t}}$ (Fig.~\ref{fig:int_exp}), we see that the expected number of events in every bin of the interference term is roughly close to half the number expected for the CP-odd signal which, in turn, is half the number of the CP-even case, both showed in Fig.~\ref{fig:angular_exp}.  

%%%%%%%%%%%%%%%. Results and Discussion
\section{Results and Discussion \label{sec:results}}
\hspace{\parindent} 

The differential angular distributions and the asymmetries are used to set confidence levels (CLs) limits on the exclusion of the SM with a contribution from a CP-mixed 125~GeV Higgs boson $\phi$, assuming the SM hypothesis as the null hypothesis. The contribution of all SM backgrounds is taken into account and normalized to the LHC luminosity, just like the signal. The CLs limits are computed both for an LHC luminosity of $L$=200~fb$^{-1}$ (corresponding roughly to the RUN 2 luminosity plus the contribution from the first year of RUN 3), and for the full luminosity ($L$=3000~fb$^{-1}$) expected at the end of the High-Luminosity phase of the LHC (HL-LHC). The CLs limits are shown as contour plots in the ($\kappa$, $\tilde{\kappa}$) 2D plane (with $\kappa = \kappa_t \cos \alpha$ and $\tilde{\kappa} = \kappa_t \sin \alpha$), which is scanned using steps of 0.1 on the values of  $\kappa$ and $\tilde{\kappa}$ in the range [-2.0, 2.0] for $L$=200~fb$^{-1}$, and [-1.5, 1.5], for $L$=3000~fb$^{-1}$.  The calculation of the CLs follows the prescription described in~\cite{Read:2002hq, Junk:1999kv}. \\[2mm]

\begin{figure}[h!]
	\begin{center}
		\hspace{-5mm}
		\includegraphics[height = 5.5cm]{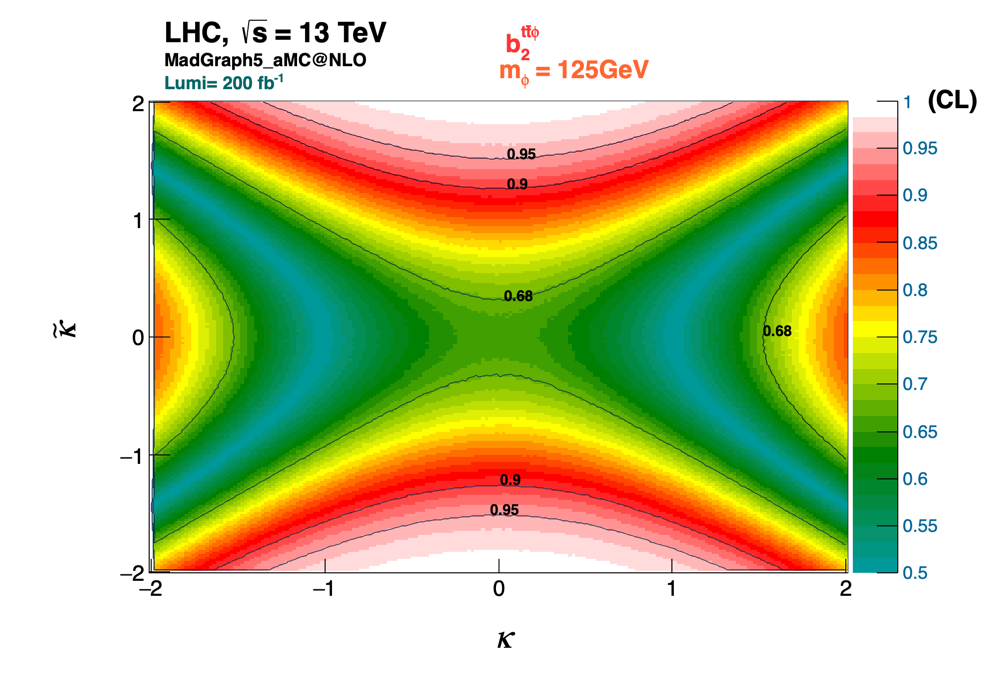}
		\hspace{-3mm}
		\includegraphics[height = 5.5cm]{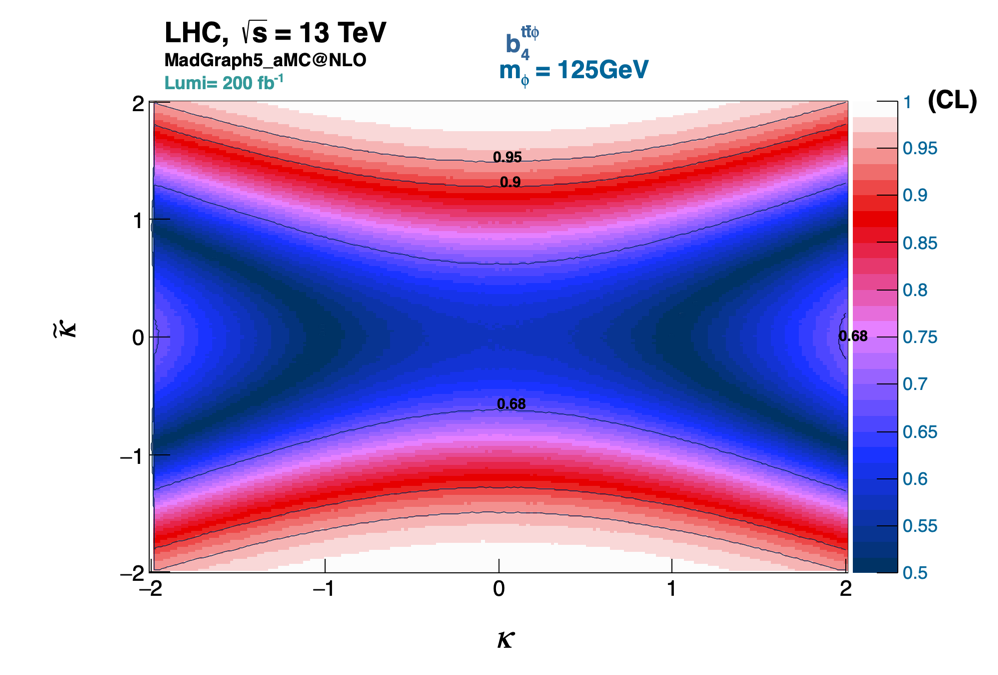}
		\\
		\hspace{-5mm}
		\includegraphics[height = 5.5cm]{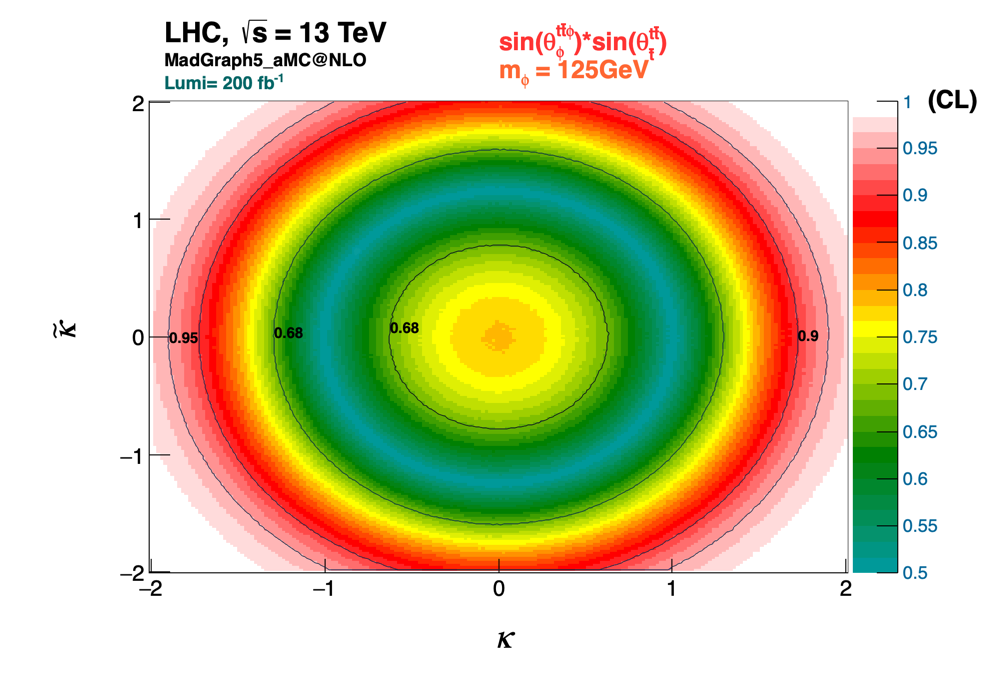}
		\hspace{-3mm}
		\includegraphics[height = 5.5cm]{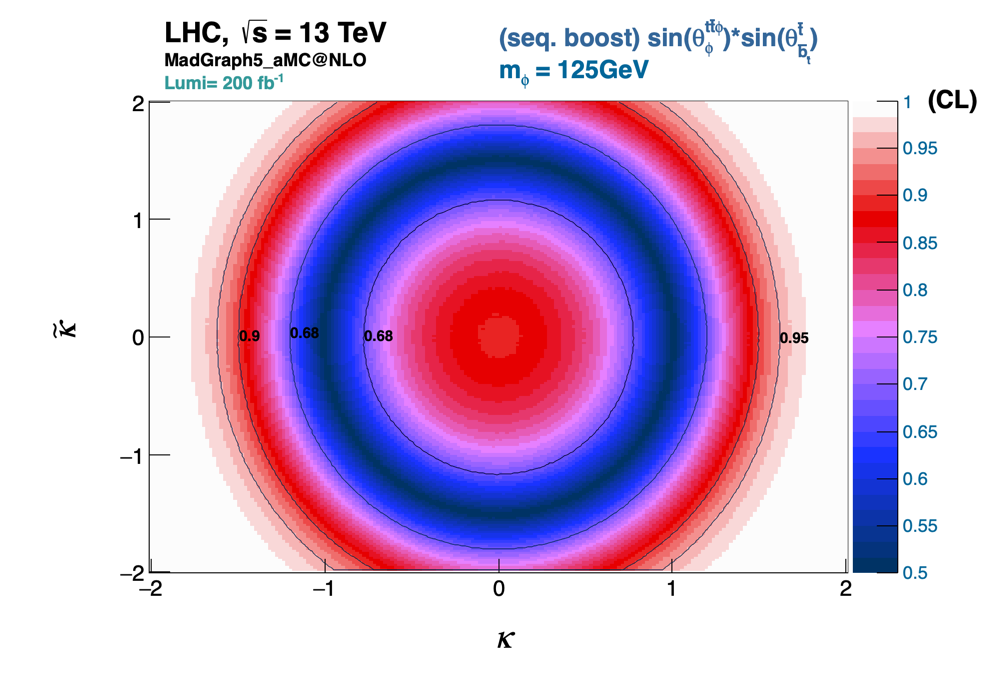}
		\caption{CLs for the exclusion of the SM with a 125~GeV Higgs boson $\phi$ with mixed scalar and pseudo-scalar couplings (CP-mixed case), against the SM as null hypothesis. Limits are shown for a luminosity corresponding to the full RUN 2 data and first year of RUN 3 i.e., $L$=200~fb$^{-1}$.}
		\label{fig:asym_200fb}
	\end{center}
\end{figure}

\noindent
{\it Exclusion Limits from Asymmetries}\\[3mm]
\noindent
Asymmetries are more interesting to explore during the initial phase of RUN 3 since they do not require the same amount of data as the differential distributions do, to be precisely measured. Whenever a total production cross section measurement is possible for the  $t\bar{t}\phi$ process at RUN 3, the evaluation of an asymmetry, i.e. a two bin measurement, is almost immediately accessible. In Fig.~\ref{fig:asym_200fb}, we have evaluated the CLs limits for the exclusion of the SM null hypothesis, by using the asymmetries introduced in this paper. As can be seen in Fig.~\ref{fig:asym_200fb}, the results obtained at the LHC are competitive with the ones obtained with differential angular distributions and may be performed during an early phase of RUN 3. Moreover the shapes of the $b_2^{t\bar{t}\phi}$ and $b_4^{t\bar{t}\phi}$ exclusion limits show a particularly different behaviour when compared with the usual $\kappa^2\sigma_\text{CP-even}+\tilde\kappa^2\sigma_\text{CP-odd}$ dependence of the exclusions obtained with cross section measurements. This relates to the fact that asymmetries (see  Eq.~\ref{equ:asymmetries}) of CP-even variables have a generic dependence with the CP-couplings of the form $A\kappa^2+B\tilde\kappa^2$ but where $A$ and $B$ are not necessarily positive anymore. They are defined as

\begin{equation}
\begin{aligned}
A \propto \int_{x_{c}}^{+1}\,d\sigma_\text{CP-even} - \int_{-1}^{x_{c}}\,d{\sigma_\text{CP-even}} \, \, \, \,  \text{and} \, \,  \, \, 
B \propto \int_{x_{c}}^{+1}\,d\sigma_\text{CP-odd} - \int_{-1}^{x_{c}}\,d{\sigma_\text{CP-odd}}, \\
\end{aligned}
\label{eq:ABdependence}
\end{equation}
and their value depends on the particular choice of the cut-off value used to define the asymmetry. An appropriate choice of $x_c$ can render $A$ and/or $B$ negative, null or positive. The choice on this paper was to set $x_c$ to maximise the difference between the asymmetries from the CP-even and CP-odd pure signal cases, at parton level (see Tab.~\ref{tab:AsymGen}).

\begin{figure}[h!]
	\begin{center}
		\hspace{-5mm}
		\includegraphics[height = 5.5cm]{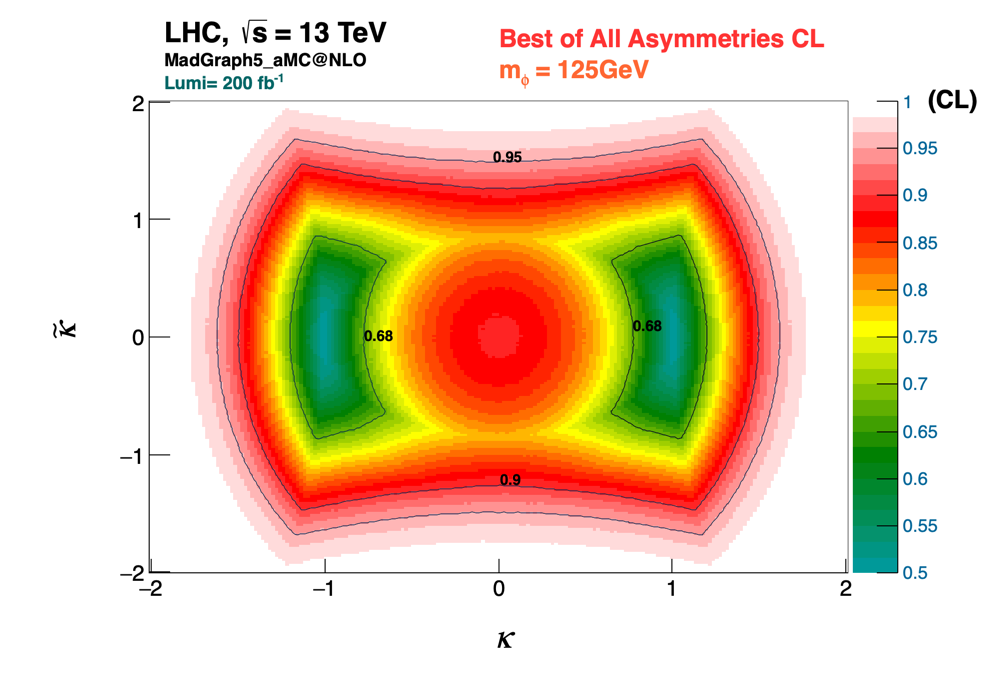}
		\hspace{-3mm}
		\includegraphics[height = 5.5cm]{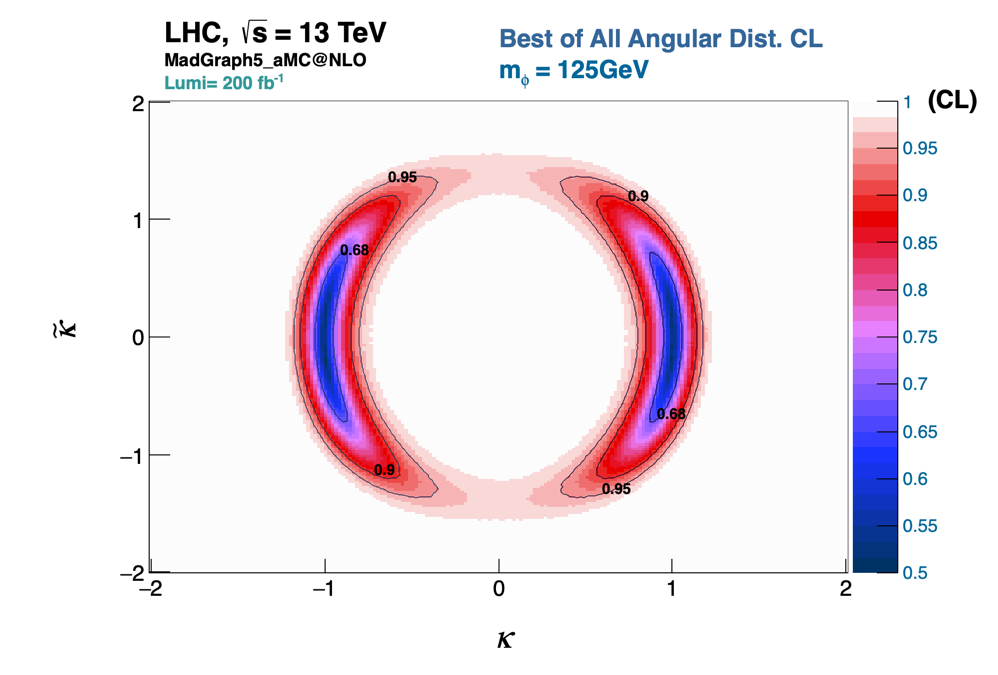}
		\caption{The best exclusion limits, from all single asymmetry exclusions, are shown (left), as well as the best exclusion limits from the angular differential distributions (right), for a luminosity corresponding to the full RUN 2 data and first year of RUN 3 i.e., $L$=200~fb$^{-1}$.}
		\label{fig:asym_best}
	\end{center}
\end{figure}

In Fig.~\ref{fig:asym_best} (left), we show the best exclusion limits obtained from the combination of the individual asymmetries in Fig.~\ref{fig:asym_200fb} where the complementarity amongst the different asymmetries becomes
clearly visible.

\clearpage

\begin{figure}[h!]
	\begin{center}
		\hspace{-5mm}
		\includegraphics[height = 5.5cm]{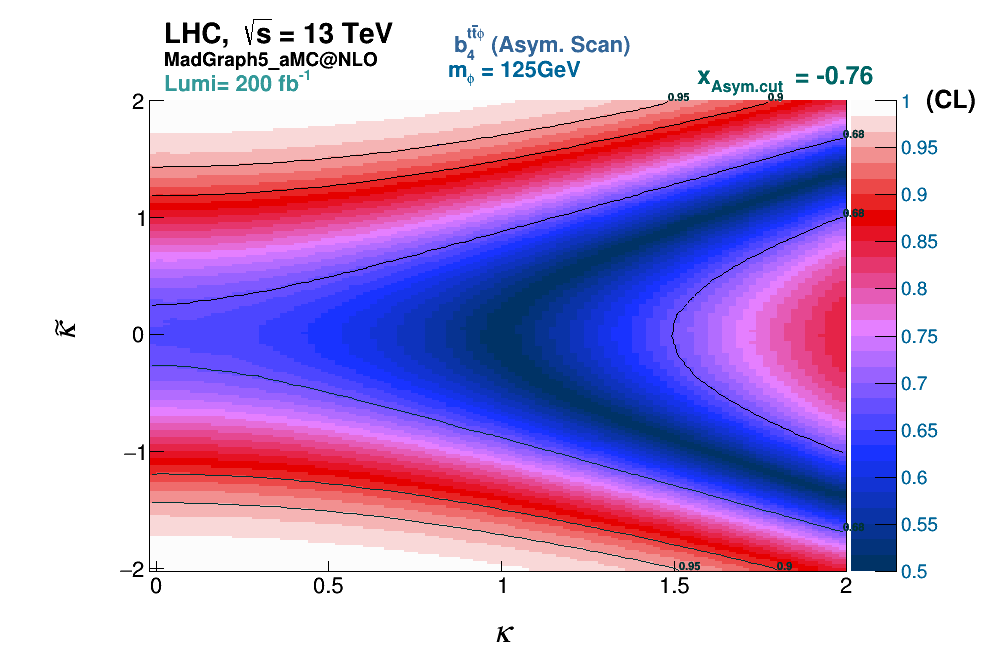}
		\hspace{-3mm}
		\includegraphics[height = 5.5cm]{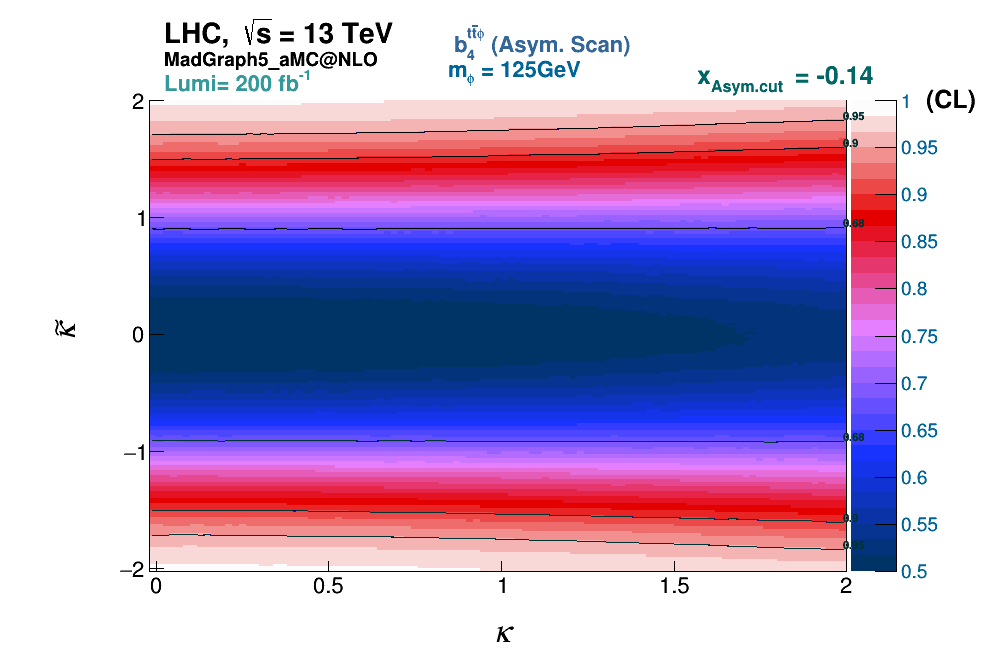}
		\\
		\hspace{-5mm}
		\includegraphics[height = 5.5cm]{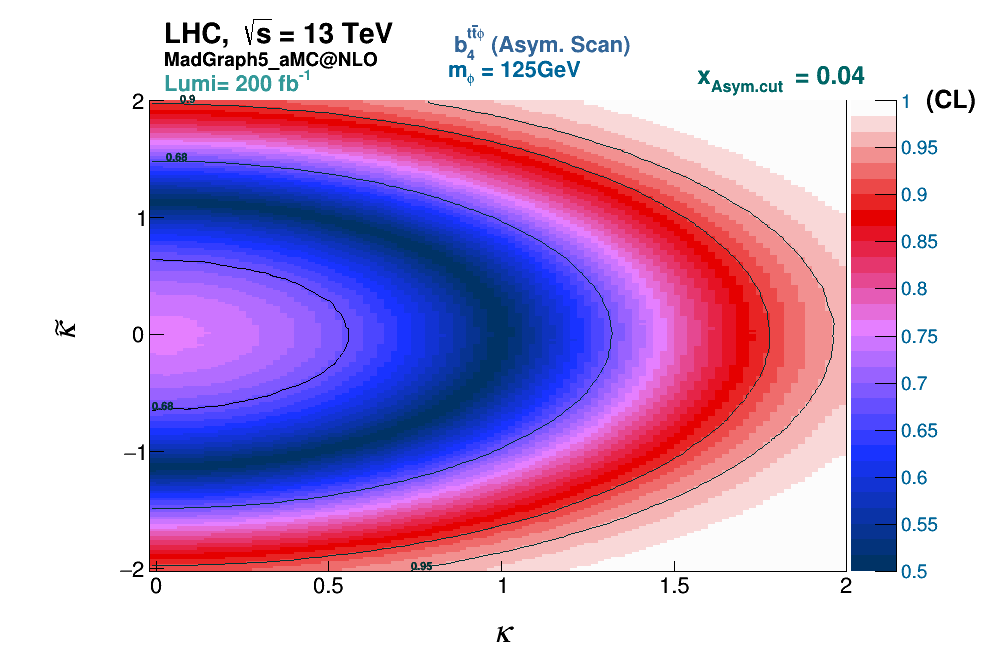}
		\hspace{-3mm}
		\includegraphics[height = 5.5cm]{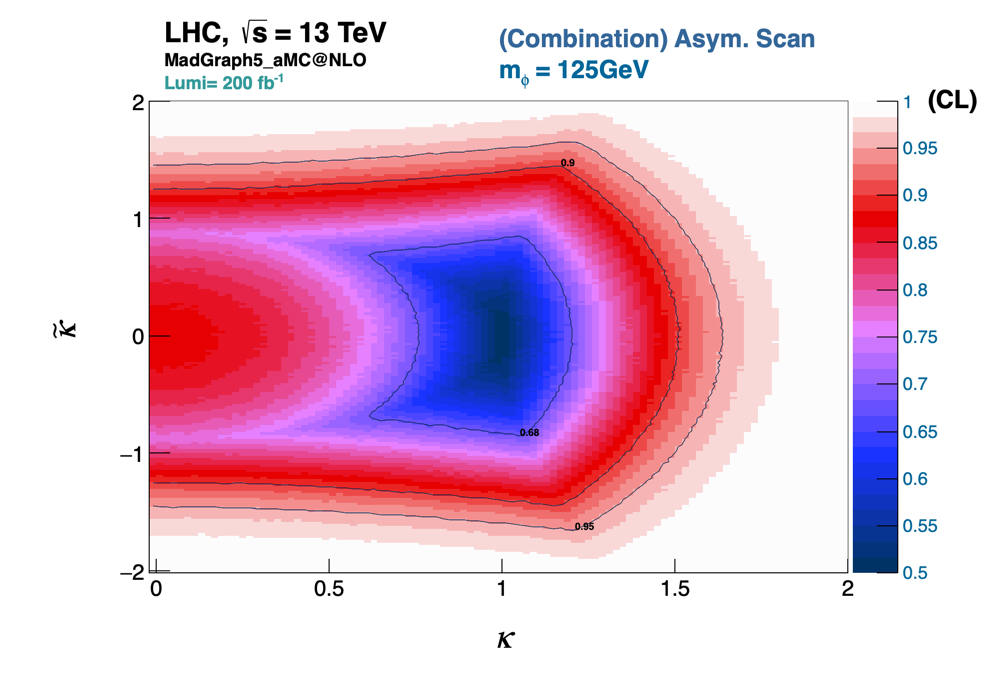}
		\caption{CLs for the exclusion of the SM with a 125~GeV Higgs boson $\phi$ with mixed scalar and pseudo-scalar couplings (CP-mixed case), against the SM as null hypothesis, for the $b_4^{t\bar{t}\phi}$ asymmetry (top- and bottom-left and top-right) and best exclusion limits from all observables (bottom-right). Limits are shown for a luminosity corresponding to the full RUN 2 data and first year of RUN 3 i.e., $L$=200~fb$^{-1}$ and as a function of different cut-off values applied to $b_4^{t\bar{t}\phi}$ i.e., $x_c$=-0.76 (top-left), $x_c$=-0.14 (top-right) and $x_c$=+0.04 (bottom-left).}
		\label{fig:scan_200fb}
	\end{center}
\end{figure}

In order to understand how a given cut-off affects the evaluation of the exclusion limits, a scan varying $x_c$ along the distributions of the four observables, defined in Eqs.~\ref{equ:b2b4} and~\ref{equ:sinsin} (and shown in Fig.~\ref{fig:asym_200fb}), was performed using 100 different values, uniformly distributed over the range of the angular distributions. It is worth mentioning that the scans, as they use distributions after event selection, kinematic reconstruction, as well a full set of SM backgrounds, allow to find the optimal asymmetry cut-off value for each observable, and get the best exclusion limits with a setup very close to a real LHC experiment, i.e. taking into account the experimental distortions and particular shape of the signals and SM backgrounds. In Fig.~\ref{fig:scan_200fb}, the $b_4^{t\bar{t}\phi}$ exclusion limits are shown for  $x_c$=-0.76 (top-left), $x_c$=-0.14 (top-right) and $x_c$=+0.04 (bottom-left), considering $k>0.0$ for simplicity. The shape of the exclusion limits change quite remarkably as a function of $x_c$. The same behaviour is observed in the other variables. It is possible to make the exclusion limits almost insensitive to the real part of the coupling (for $x_c$=-0.14), by allowing in Eq.~\ref{eq:ABdependence} the $k^2$ dependency to essentially vanish ($A\sim 0$ ). In Fig.~\ref{fig:scan_200fb} (bottom-right), we show the best combination using $x_c$ set to -0.22, +0.04,  +0.89 and +0.15 for the 
$b^{t\bar{t}\phi}_2$, 
$b^{t\bar{t}\phi}_4$, 
$\sin (\theta^{t\bar{t}\phi}_\phi)*\sin(\theta^{t\bar{t}}_{\bar{t}})$ and 
$\sin (\theta^{t\bar{t}\phi}_\phi)*\sin(\theta^{\bar{t}}_{\bar{b}_{\bar{t}}})$ (seq.), respectively. \\

By comparing the best exclusion limits obtained with the scan in Fig.~\ref{fig:scan_200fb} (bottom-right) and the limits obtained in Fig.~\ref{fig:asym_best} (left), no significant improvement is observed. Although care should be taken when considering the cut-off values for the evaluation of the individual asymmetries exclusion limits, if several are combined together the dependency with the cut-off is then largely suppressed. The simple approach of defining cut-off values that maximize the difference between the asymmetries for the pure CP-even and CP-odd cases seems appropriate for the analysis strategy. It should be stressed that no systematic uncertainties were considered in these studies and they may change the precision with which asymmetries can be measured. Nevertheless, as asymmetries are expected to be less affected by the systematic uncertainties (as they involve ratios of cross-sections where systematic uncertainties are expected to cancel out), they should be considered in an early phase of the RUN 3 of the LHC, where the signal and background modelling is not yet completely under control, providing similar exclusion limits to differential distributions. \\

\noindent
{\it Exclusion Limits from Differential Distributions}\\[3mm]
\noindent
When signal and background modelling becomes better understood at RUN 3, differential distributions will play an increasingly relevant role on setting exclusion limits. For comparison we show, in Fig.~\ref{fig:asym_best} (right), the best exclusions limits obtained with the differential angular distributions. An improvement is visible when compared with the asymmetries exclusions, see Fig.~\ref{fig:asym_best} (left). We should bear in mind, however, that the results of differential distributions may degrade once systematic uncertainties are included. This indeed motivates to explore asymmetries during the early phase of RUN 3, for a new centre-of-mass energy, when a new cross section value for $t\bar{t}\phi$ will become feasible. Both asymmetries and cross section measurements may be performed simultaneously.

\begin{figure}[H]
	\begin{center}
		\includegraphics[height = 7.0cm]{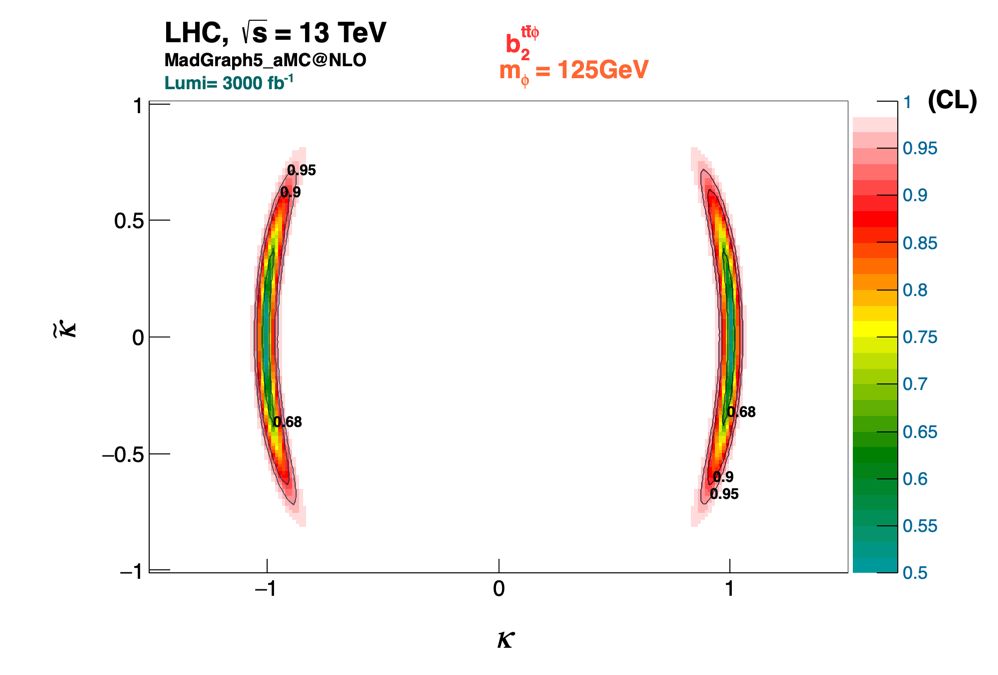}\\[1mm]
		\includegraphics[height = 7.0cm]{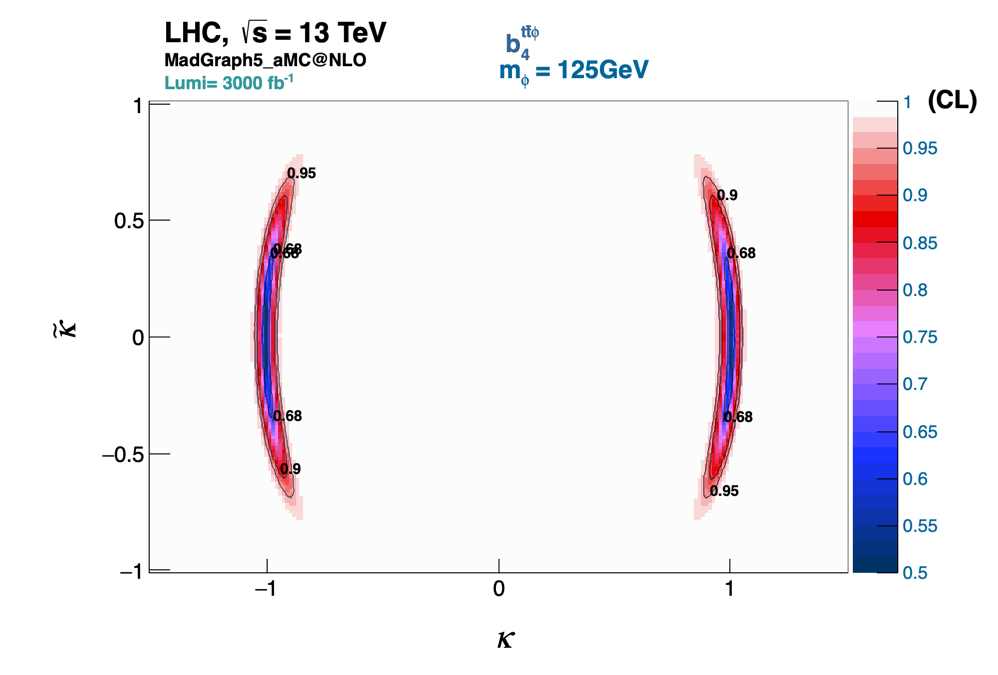}\\[1mm]
		\includegraphics[height = 7.0cm]{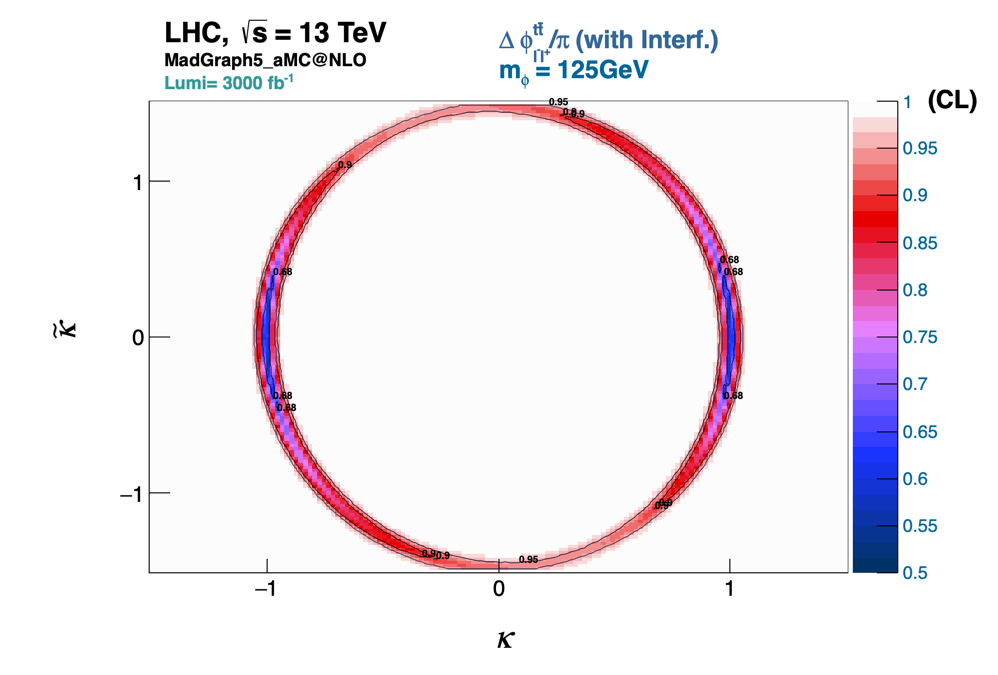}\\[1mm]
		\caption{Expected exclusion limits for the $t\bar{t}\phi$ CP-couplings for a fixed luminosity of 3000~fb$^{-1}$ and for the $b_2^{t\bar{t}\phi}$ (top) and $b_4^{t\bar{t}\phi}$ (middle) variables are shown, at the HL-LHC. The effect of interference effects is also shown (bottom) for the $\Delta \phi^{t\bar{t}}_{ll}$ angular distribution.}
		\label{fig:res3000fb}
	\end{center}
\end{figure}

In Fig.~\ref{fig:res3000fb}, we show the exclusion limits obtained when considering the full luminosity of 3000~fb$^{-1}$ at HL-LHC. The $b_2^{t\bar{t}\phi}$ (Fig.~\ref{fig:res3000fb} top) and $b_4^{t\bar{t}\phi}$ (Fig.~\ref{fig:res3000fb} middle) variables are considered as benchmark observables. In Tab.~\ref{table:exclusion_limits}, the corresponding  limits are presented at 68\% and 95\% CL. As expected, a significant improvement is observed with respect to the RUN 3 results.

\begin{table}[H]
	\renewcommand{\arraystretch}{1.3}
	\begin{center}
		\begin{tabular}{|c|c|cc|cc|}
			
			%\toprule
			\hline
			%\multirow{3}{3.5cm}{\centering $L$ = 3000~fb$^{-1}$}
			\multicolumn{2}{|c|}{}  & 
			\multicolumn{2}{c|}{Exclusion Limits} & \multicolumn{2}{|c|}{Exclusion Limits} \\ [-1mm]
			\multicolumn{2}{|c|}{$L$ = 3000~fb$^{-1}$}  & \multicolumn{2}{c|}{from $b_2^{t\bar{t}\phi}$} & \multicolumn{2}{|c|}{from $b_4^{t\bar{t}\phi}$} \\
			
			%\midrule 
			\multicolumn{2}{|c|}{}  & (68\% CL)	& (95\% CL) & (68\% CL) & (95\% CL)  \\ \hline 
			\hspace*{-3mm} \multirow{2}{2.7cm}{\centering $m_\phi$ = 125~GeV} 
			& $|\kappa| \in$ & 
			  [0.968, 1.01] & [0.878, 1.04] & [0.968, 1.01]  & [0.878, 1.04]   \\               
			& $\tilde{\kappa} \in$ & 
			  [-0.383, +0.383] & [-0.713, +0.713]  & [-0.368, +0.368] & [-0.698, +0.698]  
			\\ \hline          
			%\bottomrule
		\end{tabular}
		\caption{Expected exclusion limits for the $t\bar{t}\phi$ CP-couplings for a fixed luminosity of 3000~fb$^{-1}$. The limits are shown for 68\% and 95\% CL, for the variables $b_2^{t\bar{t}\phi}$ and $b_4^{t\bar{t}\phi}$, at the HL-LHC.}
		\label{table:exclusion_limits}
	\end{center}
\end{table}

The effect of interference terms on the exclusion limits were also studied at the HL-LHC, considering the $\Delta \phi^{t\bar{t}}_{ll}$ angular distribution, see Eq.~\ref{equ:deltaphi}. Although interference effects are visible for small values of $k$ (around zero) and large values of $\tilde{k}$ in Fig.~\ref{fig:res3000fb} (bottom), where a slight difference appears in the distribution for positive and negative values of $\tilde{k}$ (for $k \sim 0.0$), the effect is rather marginal. 

This is due to the fact that the interference terms gives a rather small contribution to the differential cross section, largely dominated by the CP-even and CP-odd terms.  By looking into the $b_2^{t\bar{t}\phi}$ or $b_4^{t\bar{t}\phi}$ exclusion limits, we see that if an exclusion is set to $|\tilde{k}|\le$~0.85, the sensitivity to the interference term vanishes completely, making the test of the interference terms in $t\bar{t}\phi$ events at the LHC almost impossible even at HL-LHC.\\

%%%%%%%%%%%%%%%. Conclusions 
\section{Conclusions \label{sec:conclusions}}
\hspace{\parindent} %forca identacao

In this paper we have studied the sensitivity to probe the CP nature of a 125~GeV Higgs boson with CP-even and CP-odd mixed couplings, at the LHC, for luminosities which typically are expected to be within reach during RUN 3 ($\sim$300~fb$^{-1}$), up to the High Luminosity phase of the LHC (HL-LHC), with 3000~fb$^{-1}$.  Signal events from $t\bar{t}\phi$ associated production are searched for, in dileptonic final states. 
While the $t\bar{t}$ system decays to two opposite charged leptons, the $\phi$ boson is expected to decay through the $\phi\rightarrow b\bar{b}$ channel. A new reconstruction method for the SM Higgs boson mass ($m_\phi$) was applied allowing to gain, in terms of mass resolution, roughly a factor of two with respect to previous analysis methods. Without loss of generality, the method can be easily extrapolated to any other two body decays of the Higgs boson like for instance $\phi\rightarrow \gamma\gamma$, provided the decay channel is kinematically accessible.

As a result of several years of testing CP-probing variables, four CP-observables,
$b^{t\bar{t}\phi}_2$, 
$b^{t\bar{t}\phi}_4$, 
$\sin (\theta^{t\bar{t}\phi}_\phi)*\sin(\theta^{t\bar{t}}_{\bar{t}})$ and 
$\sin (\theta^{t\bar{t}\phi}_\phi)*\sin(\theta^{\bar{t}}_{\bar{b}_{\bar{t}}})$ seq., were constructed and used to set exclusion limits confidence levels (CL) in the 2D ($\kappa$, $\tilde{\kappa}$) plane. A signal hypothesis with a 125~GeV Higgs boson ($\phi$) with mixed CP states, was studied against the SM scalar Higgs boson (null) hypothesis. 
The 95\% CL exclusion limits in the ($\kappa$,$\tilde{\kappa}$) plane are set to $\tilde{\kappa} \in$ [-0.698, +0.698] and $|k| \in$ [0.878, 1.04] respectively at the HL-LHC, using only the dileptonic decay channel of the $pp\rightarrow t\bar{t}\phi$ system (with $\phi\rightarrow b\bar{b}$). These results are expected to be significantly improved when the semileptonic decays of the $t\bar{t}\phi$ system are combined on one hand with the remaining decay channels of the top-quarks (W-bosons) and on the other hand with the other decay channels of the 125~GeV Higgs boson.

Exclusion limits are also presented using asymmetries built from the angular CP-observables, for specific choices of the cut-off values ($x_c$) used to define the asymmetries. The optimal values of $x_c$, i.e., the values that lead to the best exclusion limits when combining the different asymmetries, are studied. The combination of the results allowed to set exclusion limits with asymmetries almost as good as the ones obtained with differential angular distributions.  This is particularly appropriated for an early phase of the RUN 3 of the LHC, when background and signal modelling may not yet be fully under control and the statistical size of the collected data sample is still not too large. If all asymmetries are used, the cut-off can be defined by maximising the difference between the pure CP-even and CP-odd cases, at parton level, without loss of sensitivity. 

Interference terms (between CP-even and CP-odd components) on the exclusion limits, are also studied at the HL-LHC. Although interference effects are visible for small values of $k$ (around zero) and large values of $\tilde{k}$ in the ($\kappa$, $\tilde{\kappa}$) plane, the effect is rather small. In particular, if we see that an exclusion limit is set at $|\tilde{k}|\le$~0.85, the sensitivity to the interference term vanishes completely, making the test of the interference terms in $t\bar{t}\phi$ events at the LHC, unfeasible at HL-LHC.

%%%%%%%%%%%%%%%%%%%%%%%%%%%%%%%%%%%%%%%%%%%%%%%%%%%%%%%
\subsubsection*{Acknowledgments}
%\hspace{\parindent} %forca identacao
%
RC and RS are partially supported by the Portuguese Foundation for Science and Technology (FCT) under Contracts no. UIDB/00618/2020, UIDP/00618/2020, PTDC/FIS-PAR/31000/2017 and CERN/FIS-PAR/0014/2019. RC is additionally supported by FCT grant 2020.08221.BD. AO is supported by FCT under contracts no. CERN/FIS-PAR/0029/2019 and CERN/FIS-PAR/0037/2021.

%DA, RC and RS are partially supported by the Portuguese Foundation for Science and Technology (FCT) under Contracts no. UIDB/00618/2020, UIDP/00618/2020, PTDC/FIS-PAR/31000 /2017 and CERN/FIS-PAR/0002/2017, and the HARMONIA project under contract UMO-2015/18/M/ ST2/00518. AO is partially supported by FCT, under the Contract CERN/FIS-PAR/0037/2021. DA is supported by ULisboa - BD2018.

%\vspace*{0.5cm}

%%%%%%%%%%%%%%%%%%%%%%%%%%%%%%%%%%%%%%%%%%%%%%%%%%%%%%%
%%%%%%%%%%%%%%%%%%%%%%%%%%%%%%%%%%%%%%%%%%%%%%%%%%%%%%%%%%%%

\vspace*{1cm}
\bibliographystyle{h-physrev}
\bibliography{papernovo.bib}
%\bibliography{direct.bib}
%\end{comment}

\end{document}